\documentclass[11pt]{article}
\usepackage{amscd, amssymb, amsmath,amsthm}
\usepackage{yhmath,mathrsfs,xfrac}
\input{xy}
\xyoption{all}

\usepackage[a4paper,
            left=1in,
            right=1in,
            top=1in,
            bottom=1.25in,
            footskip=.5in]{geometry}
\usepackage{relsize}
\usepackage{type1cm}
\usepackage{hyperref}
\usepackage{tikz}
\usepackage{tikz-cd}
\usepackage{float}
\usepackage[utf8]{inputenc}
\usepackage{graphicx}
\usepackage{mathtools}
\usepackage{adjustbox}
\usepackage{subcaption}
\usepackage{wrapfig}
\usepackage{chemarrow}
\usepackage{leftindex}
\usepackage{comment}
\usepackage{authblk}
\usepackage[title]{appendix}
\usepackage[sort&compress,numbers]{natbib}
\numberwithin{equation}{section}
\makeatletter
\newcommand{\chapterauthor}[1]{
  {\parindent0pt\vspace*{-25pt}
  \linespread{1.1}\large\scshape#1
  \par\nobreak\vspace*{35pt}}
  \@afterheading
}
\makeatother

\theoremstyle{definition}

\DeclareMathOperator{\C}{\mathcal{C}}

\DeclareMathOperator{\Aut}{\mathrm{Aut}}
\DeclareMathOperator{\Hom}{\mathrm{Hom}_{\C}}

\DeclareMathOperator{\id}{\mathrm{id}}

\newcommand\vsim{\rotatebox[origin=cc] {90} {$ \sim $}}

\title{Lattice Translation Modulated Symmetries and TFTs}
\author[1,2]{Ching-Yu Yao}
\affil[1]{\textit{Institute for Solid State Physics, University of Tokyo, Chiba 277-8583, Japan}}
\affil[2]{\textit{Department of Physics, University of Tokyo, Tokyo 113-0033, Japan}}

\date{}
\begin{document}
\maketitle

\pagenumbering{arabic}
\setcounter{tocdepth}{2}
\setcounter{secnumdepth}{2}

Modulated symmetries are internal symmetries that are not invariant under spacetime symmetry actions. We propose a general way to describe the lattice translation modulated symmetries in 1+1D, including the non-invertible ones, via the tensor network language. We demonstrate that the modulations can be described by some autoequivalences of the categories. Although the topological behaviors are broken because of the presence of modulations, we can still construct the modulated version of the symmetry TFT bulks by inserting a series of domain walls described by invertible bimodule categories. This structure not only recovers some known results on invertible modulated symmetries but also provides a general framework to tackle modulated symmetries in a more general setting.

\tableofcontents
\newpage
\newpage
\section{Introduction}
In many-body quantum systems, symmetries play important roles in understanding physical phenomena and classifying phases of matter. The Landau paradigm distinguishes different phases by patterns of spontaneous symmetry breaking (SSB); on the other hand, the symmetry-protected topological (SPT) phases are quantum phases that do not undergo SSB, but can only be adiabatically deformed to each other if the symmetry is broken \cite{SPT 1D Spin, 1D SPT, SPT, Weak SPT, SPT Review}.

Some physical systems, such as topological orders and models with non-invertible symmetries, fail to fit the above framework; attempts to use symmetries to describe these systems lead to generalized notions of symmetries beyond ordinary internal symmetries \cite{Generalized Global Symmetries, Generalized Symmetries in Condensed Matter, ICTP, Note on Generalized Symmetries}. For a $(d+1)$-dimensional quantum system, symmetry is expected to be characterized by a fusion $(d+2)$-category \cite{Higher Categories, topological orders in any dimensions, Condensations in higher categories}. The generalized notion of symmetry with the corresponding phases can be described by a one-dimensional higher symmetry topological field theory (SymTFT) bulk \cite{Boundary-Bulk, Topological Holography, Symmetry as a shadow}. In this framework, the SymTFT bulk is described by the center of the fusion $(d+2)$-category, and the phases are realized as different choices of boundaries.

The SymTFT framework is based on the assumption that the symmetry operators are topological. In general, we should also be able to consider modulated symmetries, which are internal symmetries that are not invariant under the spacetime symmetry actions, such as multipole symmetries, exponential symmetries, and subsystem symmetries. Modulated symmetries have once again captured the attention of physicists recently because of their exotic phenomena \cite{Hilbert space fragmentation, Anomalous diffusion, Gauge Theory and Fractons, Dipole superfluid hydrodynamics, multipole-conserving diffusion}. Especially, exotic phenomena of fracton orders come from the subsystem symmetries \cite{Localization in fracton, Fracton Gauge Principle, Fracton hydrodynamics}, which are examples of modulated symmetries.

The simplest modulated symmetries are invertible modulated symmetries, which can be characterized by the group homomorphism
\begin{equation}
    \phi:G_{sp}\to\Aut(G_{int}),
\end{equation}
where $G_{int}$ is the internal symmetry group and $G_{sp}$ is the spacetime symmetry group. In this work, we are focusing on the cases where $G_{sp}=\langle T\rangle\simeq\mathbb{Z}$ is the lattice translation symmetry in 1+1D. For invertible modulated symmetries, the modulation is then given by $\phi_T:=\phi(T)\in\mathrm{Aut}(G)$. One of the features of modulated symmetries is that some of the gapped phases fail to survive under modulation. To be more specific, the uniquely gapped SPTs of a uniform symmetry $G$ in 1+1D are classified by $\psi\in H^2(G;U(1))$. But for modulated symmetries, only the cohomology classes satisfying
\begin{equation}
    \phi_T^*\psi=\psi\label{1.2}
\end{equation}
can survive with the presence of modulation. The study of modulated SPTs for some specific invertible modulated symmetries in 1+1D have been classified using matrix product states(MPS) in \cite{Dipolar SPT, Multipolar SPT}, and the general classification has been studied in \cite{Modulated SPT} using defect networks.

There have been several works on gauging the modulated symmetries, which give rise to modulated symmetries that are non-invertible \cite{NIMS, Dipole KW, Non-invertible reflection}. The motivation of this work is to generalize the study on modulated SPTs to non-invertible modulated symmetries. On the other hand, there are also some studies on the 2+1D modulated SymTFT (sometimes called spacetime symmetry-enriched SymTFT) for some specific invertible modulated symmetries \cite{Spacetime symmetry-enriched SymTFT, NIMS}. We also aim to build a realization of 2+1D modulated SymTFT for generic modulated symmetries on the 1+1D boundary. We briefly outline the contents of this paper.

In Sec. \ref{Sec2}, we consider a natural description of lattice translation modulated symmetries in 1+1D and classify the corresponding phases using the tensor network language. The construction is inspired by uniform matrix product operator (MPO) symmetries and the classification of the corresponding phases studied in \cite{MPO-SPT}. We further recover some known results for the invertible modulated symmetries and the modulated SPT. In Sec. \ref{Sec3}, we briefly review 2+1D SymTFT and propose the construction of modulated SymTFT by inserting a series of domain walls that capture the modulation on the 1+1D boundary. We build the lattice realization of the modulated SymTFT of $\mathbb{Z}_N$ dipole symmetry as an example. We also show that the continuum limit recovers the foliated BF theory \cite{2+1D Fraction, Foliated BF} for this example.
\section{Lattice Translation Modulated Symmetries in 1+1D}\label{Sec2}
In this section, we use the tensor network language to capture the category structures of lattice translation modulated symmetries and the corresponding gapped phases in 1+1D. In 1+1D, MPS and MPO are efficient tools for approximating the gapped phases \cite{MPS1, MPS2, MPS3}. The MPO symmetry and the MPO symmetry protected topological phases are classified in \cite{MPO-SPT}. To generalize the construction from uniform symmetries to modulated symmetries, the most straightforward way is to consider the MPO symmetries and the MPS gapped ground states in a site-dependent manner. We call these site-dependent MPOs and site-dependent MPSs modulated MPO and modulated MPSs.

\subsection{Modulated Symmetries}\label{Subsec2.1}
Given a uniform MPO symmetry consists of injective MPO symmetry operators. The closedness condition of fusion of the symmetry operators and the assumption of unitarity imply that the symmetry is described by a unitary fusion category \cite{MPO-SPT}. In this work, we assume that the same structure is encoded in the modulated MPO symmetry, stated as follows.

\begin{figure}[H]
\centering
\includegraphics[scale = 0.2]{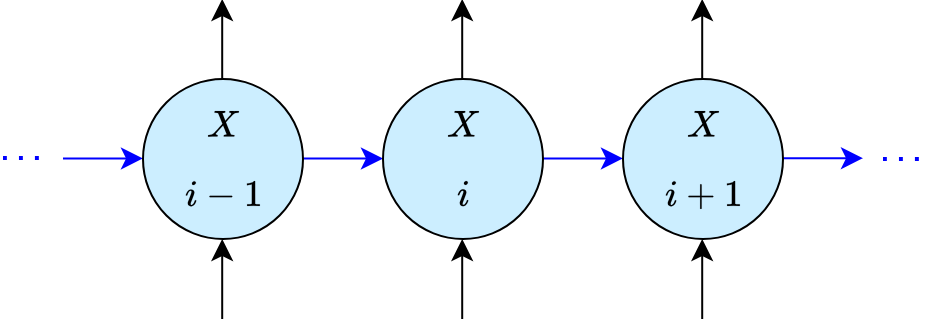}
\caption{We construct the "modulated" MPO with a site-dependent form.}
\label{MPO_1}
\end{figure}

\textbf{Symmetry Operators:} Consider the internal symmetry described by a unitary fusion category $\mathcal{C}$. Fix a set of representatives for the isomorphism classes of the simple objects in $\mathcal{C}$, denoted as $\mathcal{I}(\mathcal{C})$. The symmetry operators are injective MPOs labeled by $X\in\mathcal{I}(\mathcal{C})$, which have a site-dependent form as shown in Fig. \ref{MPO_1}.

\textbf{Fusion Rules:} 
For given $X_1,X_2,X_3\in\mathcal{I}(\mathcal{C})$, fix an orthonormal basis $B^{X_1,X_2}_{X_3}\subset\mathrm{Hom}_{\mathcal{C}}(X_1\otimes X_2,X_3)$. Since $X_1,X_2,X_3$ are simple, this implies
\begin{equation}
    X_1\otimes X_2\simeq\bigoplus_{X_3}\left|B^{X_1,X_2}_{X_3}\right|X_3,\label{2.1}
\end{equation}
which corresponds to the fusion rules of the symmetries.

Since the MPO symmetry is encoded with this $\C$-structure, we should have the decomposition
\begin{equation}
    \adjincludegraphics[valign=c, scale=0.20]{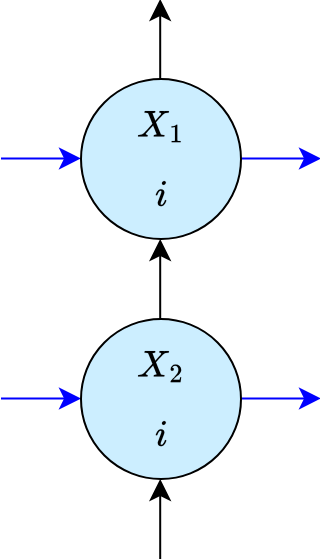}=\sum_{X_3}\sum_{\alpha}\adjincludegraphics[valign=c, scale=0.32]{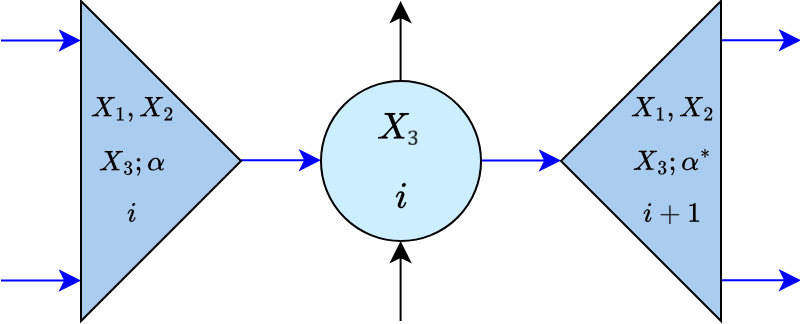},\label{2.2}
\end{equation}
summing over $X_3\in\mathcal{I}(\mathcal{C})$ and $\alpha\in B^{X_1,X_2}_{X_3}$, where the triangle gates are isometries satisfying the condition
\begin{equation}
    \adjincludegraphics[valign=c, scale=0.20]{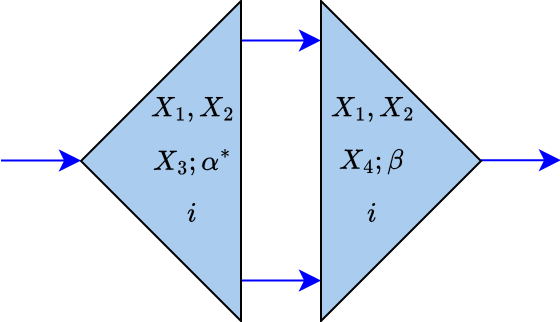}=\delta_{X_3,X_4}\delta_{\alpha,\beta}\adjincludegraphics[valign=c, scale=0.20]{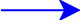}.\label{2.3}
\end{equation}
Note that the ${}^*$-structure is a contravariant endofunctor of $\C$
\begin{align}
    \begin{split}
        \Hom(X_1,X_2)&\underset{\sim}{\overset{*}{\to}}\Hom(X_2,X_1)\\
        f&\mapsto f^*
    \end{split}\label{2.4}
\end{align}
such that ${}^*\circ{}^*=\id_{\C}$.

\textbf{Associativity Natural Isomorphism:} Given $X_1,X_2,X_3,X_4\in\mathcal{I}(\mathcal{C})$, the natural isomorphism
\begin{equation}
    a_{X_1,X_2,X_3}:(X_1\otimes X_2)\otimes X_3\overset{\sim}{\to }X_1\otimes(X_2\otimes X_3)\label{2.5}
\end{equation}
defines another isomorphism
\begin{equation}
    \begin{tikzcd}
    \Hom((X_1\otimes X_2)\otimes X_3,X_4) \arrow[r,"\sim"] \arrow[d,"\vsim"]&\bigoplus_{X_5\in\mathcal{I}(\mathcal{C})}\Hom(X_1\otimes X_2,X_5)\otimes\Hom(X_5\otimes X_3,X_4) \arrow[d, "F^{X_1,X_2,X_3}_{X_4}"]\\
    \Hom(X_1\otimes(X_2\otimes X_3),X_4) \arrow[r,"\sim"] &\bigoplus_{X_6\in\mathcal{I}(\mathcal{C})}\Hom(X_2\otimes X_3,X_6)\otimes\Hom(X_1\otimes X_6,X_4)
\end{tikzcd}\label{2.6}
\end{equation}
which further defines the $F$-symbols (or $6j$-symbols) as the change of basis
\begin{equation}
    F^{X_1,X_2,X_3}_{X_4}(\alpha\otimes\beta):=\sum_{X_6}\sum_{\gamma, \delta}\left(F^{X_1,X_2,X_3}_{X_4}\right)^{X_5;\alpha,\beta}_{X_6;\gamma,\delta}\gamma\otimes\delta\label{2.7}
\end{equation}
for $X_5\in\mathcal{I}(\mathcal{C})$, $\alpha\in B^{X_1,X_2}_{X_5}$, $\beta\in B^{X_5,X_3}_{X_4}$, and summing over $X_6\in\mathcal{I}(\mathcal{C})$, $\gamma\in B^{X_2,X_3}_{X_6}$, $\delta\in B^{X_1,X_6}_{X_4}$.

The tensor network representation of the $F$-symbols is
\begin{equation}
    \adjincludegraphics[valign=c, scale=0.20]{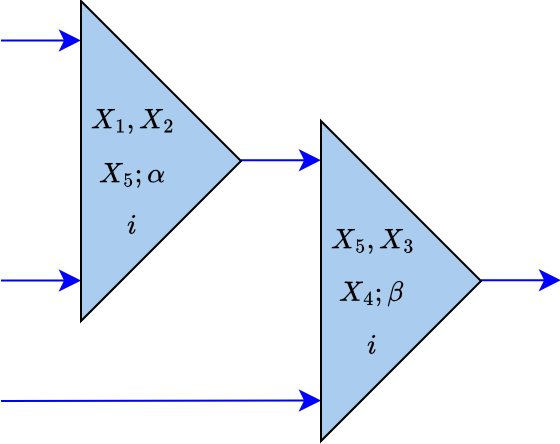}=\sum_{X_6}\sum_{\gamma,\delta}\left(F^{X_1,X_2,X_3}_{X_4}\right)^{X_5;\alpha,\beta}_{X_6;\gamma,\delta}\adjincludegraphics[valign=c, scale=0.20]{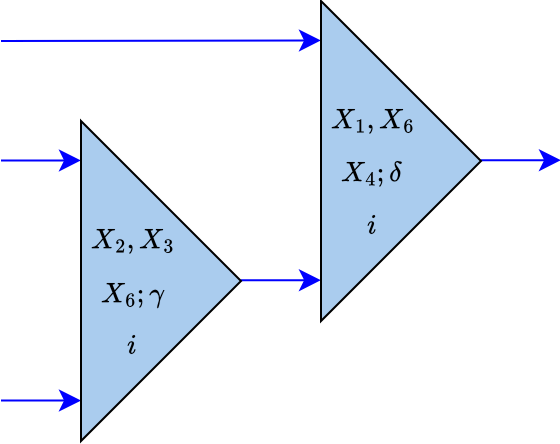},\label{2.8}
\end{equation}
which is derived from the associativity of the fusion of MPOs \cite{MPO-SPT}. Note that we assume that it is independent to $i$. We also get
\begin{equation}
    \adjincludegraphics[valign=c, scale=0.20]{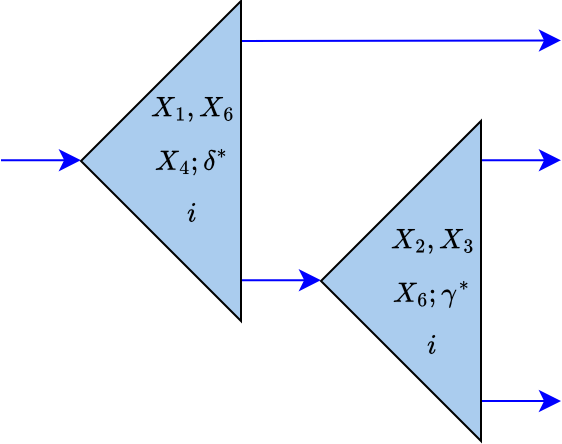}=\sum_{X_5}\sum_{\alpha,\beta}\left(F^{X_1,X_2,X_3}_{X_4}\right)^{X_5;\alpha,\beta}_{X_6;\gamma,\delta}\adjincludegraphics[valign=c, scale=0.20]{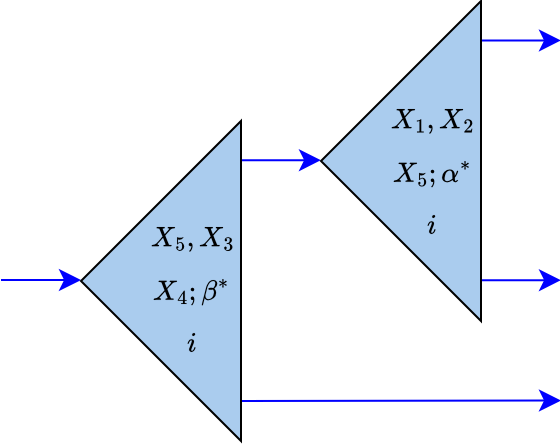}.\label{2.9}
\end{equation}

\textbf{Modulation:} In order to describe the modulation, we need to study how the symmetry operators transformed under lattice translation. It turns out that the closedness condition of the lattice translation implies that the modulation is described by a monoidal autoequivalence $F_T:\mathcal{C}\to\mathcal{C}$. The justification is as follows.

Given a symmetry action $X\in\mathcal{I}(\mathcal{C})$, it is sent to a new symmetry action $X'\in\mathcal{I}(\mathcal{C})$ under the lattice translation. This means that the $i+1$-th gate of $X$ is the $i$-th gate of $X'$ up to some unitary gates. We call the $i+1$-th gate of $X$ as the $i$-th gate of $F_TX$
\begin{equation}
    \adjincludegraphics[valign=c, scale=0.20]{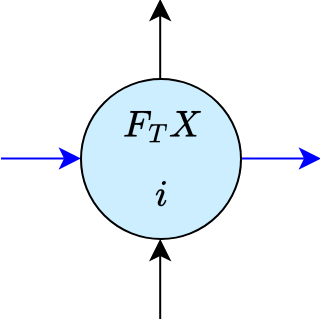}:=\adjincludegraphics[valign=c, scale=0.20]{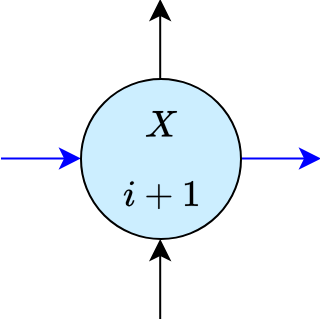}=\adjincludegraphics[valign=c, scale=0.20]{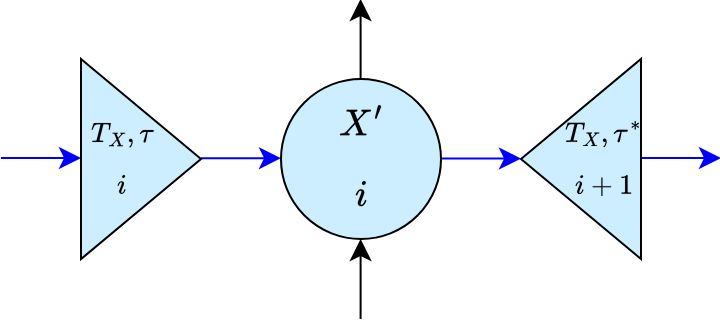},\label{2.10}
\end{equation}
$F_TX$ is then an object in $\mathcal{C}$ satisfying $F_TX\simeq X'$, and $\tau$ is labeled by a unitary vector in the 1-dimensional vector space $\Hom(F_TX,X')$.

By definition, $F_T$ is an endofunctor of $\mathcal{C}$ such that there exists a natural isomorphism
\begin{equation}
    (\eta_T)_{X_1,X_2}:F_T(X_1\otimes X_2)\overset{\sim}{\to}F_T X_1\otimes F_T X_2.\label{2.11}
\end{equation}
Furthermore, we can construct another functor $F_{T^{-1}}$ with respect to the inverse lattice translation $T^{-1}$ in the same way. $F_T\circ F_{T^{-1}}$ and $F_{T^{-1}}\circ F_T$ are equivalent to the identity functor $\id_\mathcal{C}$ by definition. Hence, $F_T$ is an autoequivalence. Note that an autoequivalence always sends simple objects to simple objects, which is a property we expect to have in the above construction.

(\ref{2.11}) defines another isomorphism
\begin{equation}
    \begin{tikzcd}
    \Hom(F_T(X_1\otimes X_2),X_3') \arrow[r,"\sim"] \arrow[d,"\vsim"]&\Hom(X_1\otimes X_2,X_3)\otimes\Hom(F_TX_3,X_3') \arrow[d, "(\eta_T)^{X_1,X_2}_{X_3}"]\\
    \Hom(F_TX_1\otimes F_TX_2,X_3) \arrow[r,"\sim"] &\Hom(F_TX_1,X_1')\otimes\Hom(F_TX_2,X_2')\otimes\Hom(X_1'\otimes X_2',X_3')\label{2.12}
\end{tikzcd}
\end{equation}
which further defines the change of basis
\begin{equation}
    (\eta_T)^{X_1,X_2}_{X_3}(\alpha\otimes\tau_3):=\sum_{\alpha'}\left((\eta_T)^{X_1,X_2}_{X_3}\right)^{\alpha,\tau_3}_{\alpha',\tau_1,\tau_2} \tau_1\otimes\tau_2\otimes\alpha'\label{2.13}
\end{equation}
for $\alpha\in B^{X_1,X_2}_{X_3}$, $\alpha'\in B^{X_1',X_2'}_{X_3'}$, and unitary $\tau_a\in\Hom(F_TX_a,X_a')$ ($a=1,2,3$).

(\ref{2.13}) is captured by the tensor networks as follows. Fusing the $i+1$-th gates of $X_1$ and $X_2$, we have the equation
\begin{align}
    \begin{split}
        &\sum_{X_3}\sum_{\alpha}\adjincludegraphics[valign=c, scale=0.20]{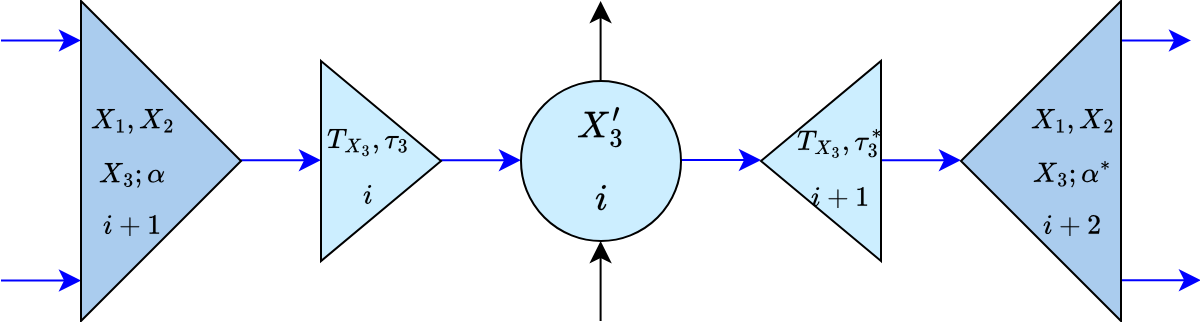}\\
        =&\sum_{X_3}\sum_{\alpha'}\adjincludegraphics[valign=c, scale=0.20]{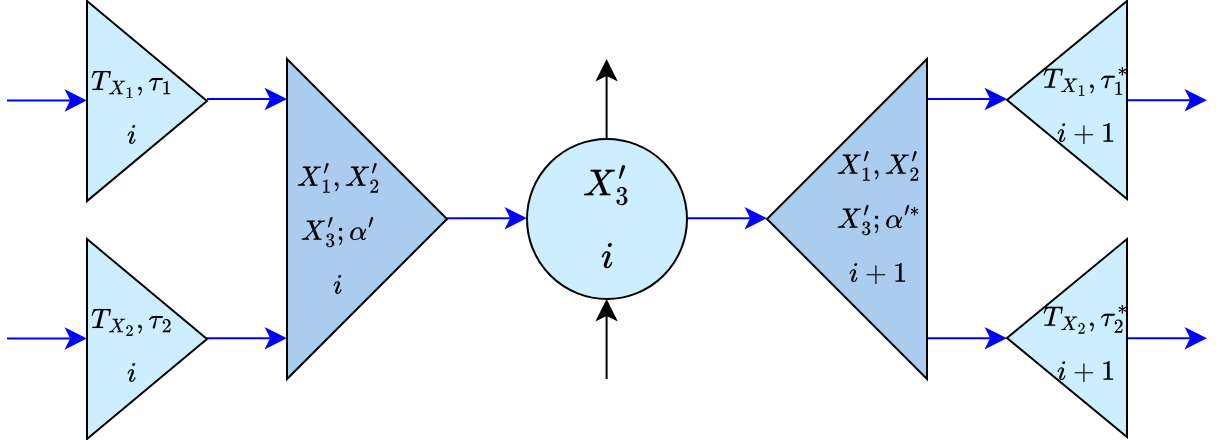}.
    \end{split}\label{2.14}
\end{align}
Define
\begin{equation}
    \left((\eta_T)^{X_1,X_2}_{X_3}\right)^{\alpha,\tau_3}_{\alpha',\tau_1,\tau_2}\adjincludegraphics[valign=c, scale=0.20]{2/MPO_3_2.png}:=\adjincludegraphics[valign=c, scale=0.20]{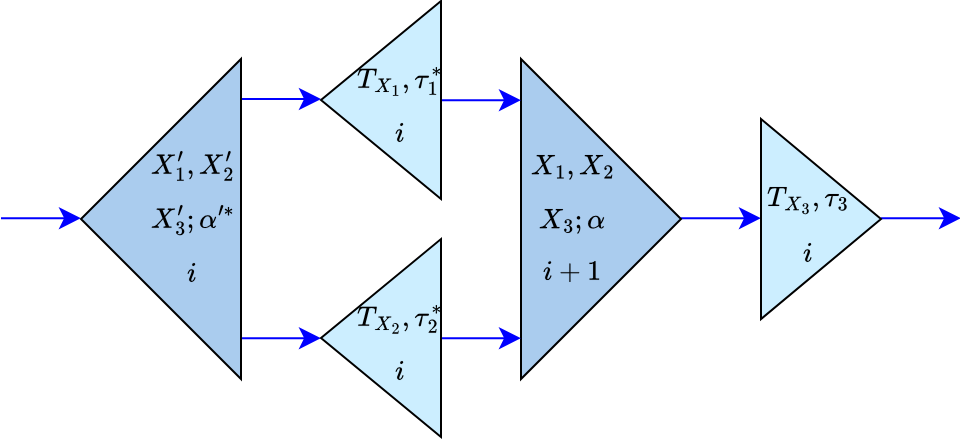},\label{2.15}
\end{equation}
and assume that it is independent to $i$. Then, due to (\ref{2.3}) and the injectivity of these MPOs, $\left((\eta_T)^{X_1,X_2}_{X_3}\right)^{\alpha,\tau_3}_{\alpha',\tau_1,\tau_2}$ is a complex number.

Lastly, according to (\ref{2.8}), (\ref{2.9}), and (\ref{2.15}), we have
\begin{align}
    \begin{split}
        &\adjincludegraphics[valign=c, scale=0.20]{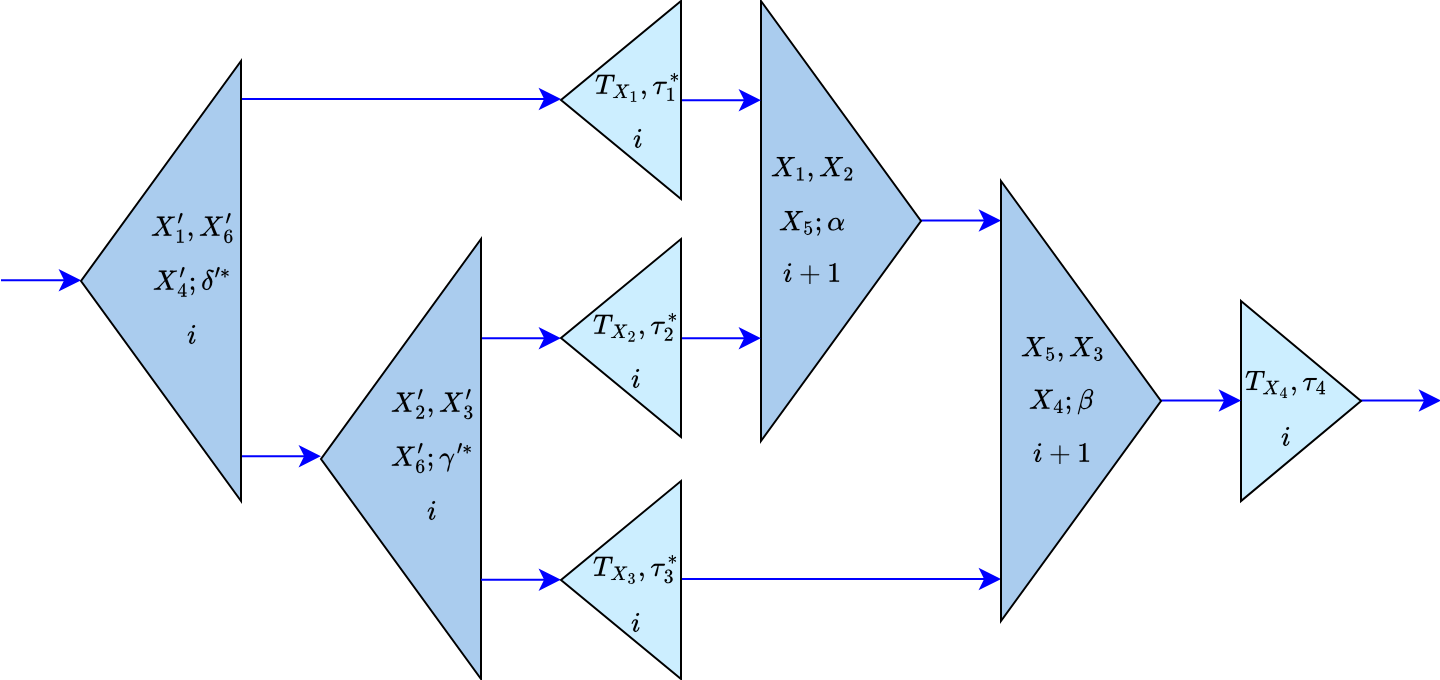}\\
        =&\sum_{X_6}\sum_{\gamma,\delta}\left((\eta_T)^{X_2,X_3}_{X_6}\right)^{\gamma,\tau_6}_{\gamma',\tau_2,\tau_3}\left((\eta_T)^{X_1,X_6}_{X_4}\right)^{\delta,\tau_4}_{\delta',\tau_1,\tau_6}\left(F^{X_1,X_2,X_3}_{X_4}\right)^{X_5;\alpha,\beta}_{X_6;\gamma,\delta}\adjincludegraphics[valign=c, scale=0.20]{2/MPO_3_2.png}\\
        =&\sum_{X_5}\sum_{\alpha',\beta'}\left(F^{X_1',X_2',X_3'}_{X_4'}\right)^{X_5';\alpha',\beta'}_{X_6';\gamma',\delta'}\left((\eta_T)^{X_1,X_2}_{X_5}\right)^{\alpha,\tau_5}_{\alpha',\tau_1,\tau_2}\left((\eta_T)^{X_5,X_3}_{X_4}\right)^{\beta,\tau_4}_{\beta',\tau_5,\tau_3}\adjincludegraphics[valign=c, scale=0.20]{2/MPO_3_2.png}.\label{2.16}
    \end{split}
\end{align}
This implies that the following diagram of isomorphisms commutes
\begin{equation}
    \begin{tikzcd}
    &{F_T(X_1\otimes X_2)\otimes F_TX_3}&\\
	{F_T((X_1\otimes X_2)\otimes X_3)} && {(F_TX_1\otimes F_TX_2)\otimes F_TX_3} \\
	&& {} \\
	{F_T(X_1\otimes(X_2\otimes X_3))} &&  {F_TX_1\otimes(F_TX_2\otimes F_TX_3)}\\
    &{F_TX_1\otimes F_T(X_2\otimes X_3)}&
	\arrow["{(\eta_T)_{X_1\otimes X_2, X_3}}", from=2-1, to=1-2]
	\arrow["{F_T(a_{X_1,X_2,X_3})}"', from=2-1, to=4-1]
	\arrow["{(\eta_T)_{X_1,X_2}\otimes\mathrm{id}}", from=1-2, to=2-3]
	\arrow["{a_{F_TX_1,F_TX_2,F_TX_3}}", from=2-3, to=4-3]
	\arrow["{(\eta_T)_{X_1,X_2\otimes X_3}}"', from=4-1, to=5-2]
	\arrow["{\mathrm{id}\otimes(\eta_T)_{X_2,X_3}}"', from=5-2, to=4-3]
    \end{tikzcd}\label{2.17}
\end{equation}
Thus, $F_T$ preserves the monoidal structure. This concludes that $F_T$ is a monoidal autoequivalence of $\mathcal{C}$.

\subsection{Mixed Anomalies}\label{Subsec2.2}
Consider invertible symmetries as the simplest example. The internal $G$ symmetry is described by the category of finite-dimensional $G$-graded $\mathbb{C}$-vector spaces. Denote the fixed representatives of the isomorphism classes of simple objects as $g\in G$, and define
\begin{equation}
    \omega(g_1,g_2,g_3):=\left(F^{g_1,g_2,g_3}_{g_1g_2g_3}\right)^{g_1g_2;\alpha,\beta}_{g_2g_3;\gamma,\delta}\in U(1)
\end{equation}
for fixed $B^{g,h}_{gh}$ ($g,h\in G$). The pentagon axiom gives the condition
\begin{equation}
    d\omega=0\in C^4(G;U(1)),
\end{equation}
and the redundancy of choosing $B^{g,h}_{gh}$ gives the redundancy
\begin{equation}
    \omega\sim\omega+d\beta\label{2.20}
\end{equation}
for some $\beta\in C^2(G;U(1))$.

Now consider the modulation $F_T$. Write $\phi_T(g)\simeq F_Tg$, then $F_T(g_1\otimes g_2)\simeq F_Tg_1\otimes F_Tg_2$ implies that $\phi_T$ is a group homomorphism, and $F_T$ being an autoequivalence implies that $\phi_T\in \mathrm{Aut}(G)$. This is what we expect an invertible lattice translation modulated symmetry should look like: a group homomorphism
\begin{align}
    \begin{split}
        \phi:\mathbb{Z}&\to\mathrm{Aut}(G).\\
        n&\mapsto \left(\phi_T\right)^n
    \end{split}
\end{align}

We also have to keep track and classify the natural isomorphism $\eta_T$ of the autoequivalence. Define 
\begin{equation}
    \alpha(g_1,g_2):=\left((\eta_T)^{g_1,g_2}_{g_3}\right)^{\alpha,\tau_3}_{\alpha',\tau_1,\tau_2}
\end{equation}
for fixed $\tau_a:F_Tg\to\phi_T(g)$ ($a=1,2,3$). (\ref{2.16}) then gives the condition
\begin{equation}
    d\alpha=\phi_T^*\omega-\omega,\label{2.23}
\end{equation}
this implies that
\begin{equation}
    \phi_T^*[\omega]=[\omega]\in H^3(G;U(1)).\label{2.24}
\end{equation}
The redundancy of choosing $\tau$'s gives the redundancy
\begin{equation}
    \alpha\sim\alpha+\phi_T^*\beta-\beta+d\gamma\label{2.25}
\end{equation}
for some $\gamma\in C^1(G;U(1))$, and $\beta\in C^2(G;U(1))$ is the same as the one in (\ref{2.20}) since it comes from the redundancy of $B^{g,h}_{gh}$.

We want to classify the pairs $(\alpha,\omega)$ up to redundancy for a fixed $\phi_T\in\mathrm{Aut}(G)$. According to (\ref{2.24}), we have
\begin{equation}
    [\omega]\in H^3(G;U(1))^{\phi_T^*}:=\mathrm{ker}\left(1-\phi_T^*:H^3(G;U(1))\to H^3(G;U(1))\right).
\end{equation}
For a fixed $[\omega]$, according to (\ref{2.25}), the pairs $(\alpha,\omega)$ up to redundancy form a $H^2(G;U(1))_{\phi_T^*}$-torsor, where
\begin{equation}
    H^2(G;U(1))_{\phi_T^*}:=\mathrm{coker}\left(1-\phi_T^*:H^2(G;U(1))\to H^2(G;U(1))\right).
\end{equation}
Therefore, the classification is described by the extension of $H^3(G;U(1))^{\phi_T^*}$ by $H^2(G;U(1))_{\phi_T^*}$.

Consider the case with no modulation, i.e., $\phi_T=\mathrm{id}_G$, then (\ref{2.23}) becomes 2-cocycle condition. Thus, the extension splits, and the classification is simply $H^2(G;U(1))\oplus H^3(G;U(1))$. The cohomology class $[\alpha]\in H^2(G;U(1))$ actually captures the mixed anomaly of the internal $G$ symmetry and the lattice translation symmetry \cite{anomaly
matching, LSM}, since it corresponds to the natural isomorphism $\eta_T$, which describes the non-trivial $U(1)$ phases between changing the order of fusion symmetry operators and lattice translation.

For generic $\phi_T\in\mathrm{Aut}(G)$, the classification is described by the extension
\begin{equation}
    1\to H^2(G;U(1))_{\phi_T^*}\to H^3(G\underset{\phi}{\rtimes}\mathbb{Z};U(1))\to H^3(G;U(1))^{\phi_T^*}\to1.
\end{equation}
For detailed justification, please refer to Appendix \ref{A}.
\subsection{Modulated Symmetry Protected Topological Phases}\label{Subsec2.3}
The gapped groundstates in 1+1D systems are described by injective MPSs. Given a uniform MPO symmetry described by a fusion category $\mathcal{C}$, the corresponding symmetry-protected topological phases are classified by the indecomposable semisimple $\mathcal{C}$-module categories \cite{MPO-SPT, Gapped Phases}. In this work, we assume that the same structure is encoded in the modulated MPS groundstates, stated as follows.

\begin{figure}[H]
\centering
\includegraphics[scale = 0.18]{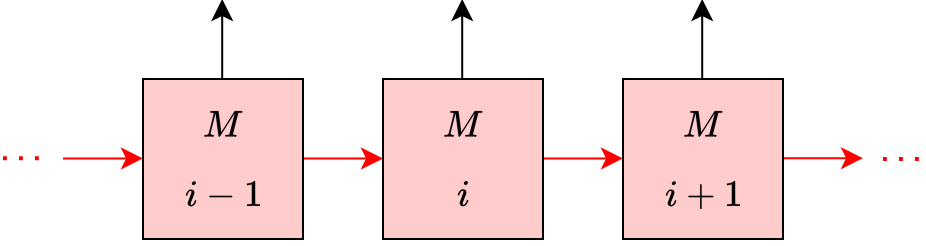}
\caption{We construct the "modulated" MPS with a site-dependent form.}
\label{MPS_1}
\end{figure}

\textbf{Ground States:} Consider a gapped phase described by an indecomposable semisimple $\mathcal{C}$-module category $\mathcal{M}$. Fix a set of representatives for the isomorphism classes of the simple objects in $\mathcal{M}$, denoted as $\mathcal{I}(\mathcal{M})$. The gapped ground states are described by modulated injective MPS, and labeled by $M\in\mathcal{I}(\mathcal{M})$, which have a site-dependent form as shown in Fig. \ref{MPS_1}

\textbf{Symmetry Actions:} The symmetry actions are described by the $\mathcal{C}$ action on $\mathcal{M}$, which is a bifunctor
\begin{align}
    \begin{split}
        \triangleright:\mathcal{C}\times\mathcal{M}&\to\mathcal{M}.\\
        (X,M)&\mapsto X\triangleright M
    \end{split}\label{2.29}
\end{align}

For given $X\in\mathcal{I}(\mathcal{C})$ and $M_1,M_2\in\mathcal{I}(\mathcal{M})$, fix an orthonormal basis $B^{X,M_1}_{M_2}\subset\mathrm{Hom}_{\mathcal{M}}(X\triangleright M_1,M_2)$. Since $X,M_1,M_2$ are simple, this implies
\begin{equation}
    X\triangleright M_1\simeq\bigoplus_{M_2}\left|B^{X,M_1}_{M_2}\right|M_2.\label{2.30}
\end{equation}

Since the MPS are encoded with this $\mathcal{M}$-structure, we should have the decomposition
\begin{equation}
    \adjincludegraphics[valign=c, scale=0.20]{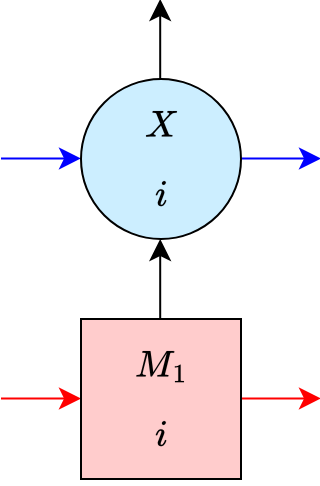}=\sum_{M_2}\sum_{\alpha}\adjincludegraphics[valign=c, scale=0.20]{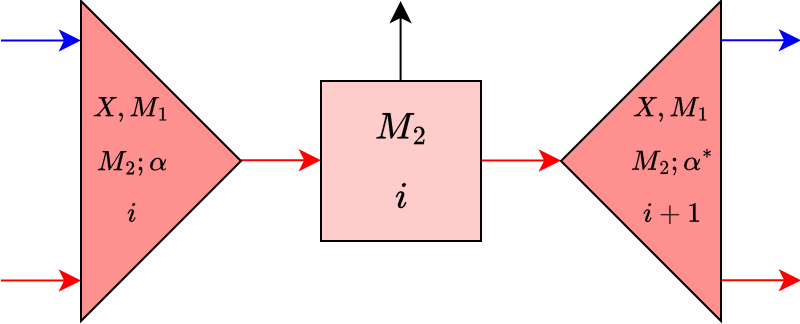},\label{2.31}
\end{equation}
summing over $M_2\in\mathcal{I}(\mathcal{M})$ and $\alpha\in B^{X,M_1}_{M_2}$, where the triangle gates are isometries satisfying the condition
\begin{equation}
    \adjincludegraphics[valign=c, scale=0.20]{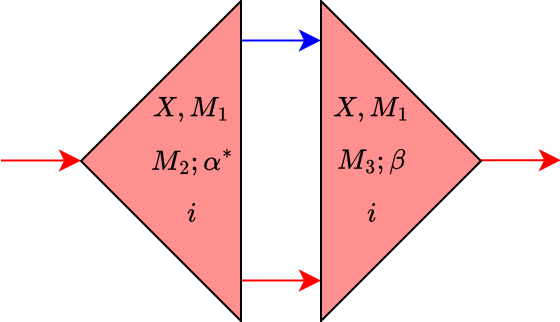}=\delta_{M_2,M_3}\delta_{\alpha,\beta}\adjincludegraphics[valign=c, scale=0.20]{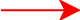}.\label{2.32}
\end{equation}
Note that the ${}^*$-structure is a contravariant endofunctor of $\mathcal{M}$
\begin{align}
    \begin{split}
        \mathrm{Hom}_\mathcal{M}(M_1,M_2)&\underset{\sim}{\overset{*}{\to}}\mathrm{Hom}_\mathcal{M}(M_2,M_1)\\
        f&\mapsto f^*
    \end{split}\label{2.33}
\end{align}
such that ${}^*\circ{}^*=\id_{\mathcal{M}}$.

\textbf{Associativity Natural Isomorphism:} Given $X_1,X_2\in\mathcal{I}(\C)$ and $M_1,M_2\in\mathcal{I}(\mathcal{M})$, the natural isomorphism
\begin{equation}
    m_{X_1,X_2,M_1}:(X_1\otimes X_2)\triangleright M_1\overset{\sim}{\to }X_1\triangleright(X_2\triangleright M_1)\label{2.34}
\end{equation}
defines another isomorphism
\begin{equation}
    \begin{tikzcd}
    \mathrm{Hom}_{\mathcal{M}}((X_1\otimes X_2)\triangleright M_1,M_2) \arrow[r,"\sim"] \arrow[d,"\vsim"]&\bigoplus_{X_3}\Hom(X_1\otimes X_2,X_3)\otimes\mathrm{Hom}_{\mathcal{M}}(X_3\triangleright M_1,M_2) \arrow[d, "^\triangleright F^{X_1,X_2,M_1}_{M_2}"]\\
    \mathrm{Hom}_{\mathcal{M}}(X_1\otimes(X_2\triangleright M_1),M_2) \arrow[r,"\sim"] &\bigoplus_{M_3}\mathrm{Hom}_{\mathcal{M}}(X_2\triangleright M_1,M_3)\otimes\mathrm{Hom}_{\mathcal{M}}(X_1\triangleright M_3,M_2)
    \end{tikzcd}\label{2.35}
\end{equation}
which further defines the ${}^{\triangleright}F$-symbols as the change of basis
\begin{equation}
    ^\triangleright F^{X_1,X_2,M_1}_{M_2}(\alpha\otimes\beta):=\sum_{M_3}\sum_{\gamma, \delta}\left(^\triangleright F^{X_1,X_2,M_1}_{M_2}\right)^{X_3;\alpha,\beta}_{M_3;\gamma,\delta}\gamma\otimes\delta.\label{2.36}
\end{equation}
for $X_3\in\mathcal{I}(\mathcal{C})$, $\alpha\in B^{X_1,X_2}_{X_3}$, $\beta\in B^{X_3,M_1}_{M_2}$, and summing over $M_3\in\mathcal{I}(\mathcal{M})$, $\gamma\in B^{X_2,M_1}_{M_3}$, $\delta\in B^{X_1,M_3}_{M_2}$. The tensor network representation of the ${}^\triangleright F$-symbols is
\begin{equation}
    \adjincludegraphics[valign=c, scale=0.20]{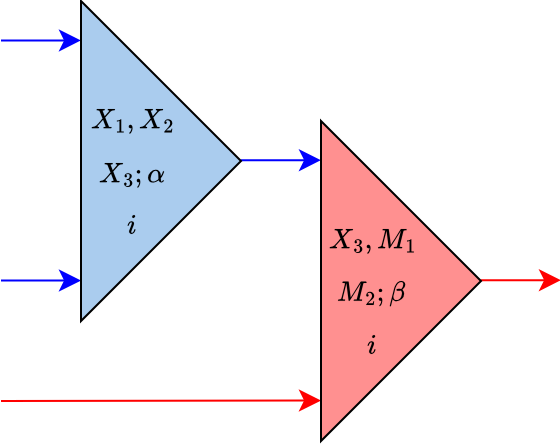}=\sum_{M_3}\sum_{\gamma,\delta}\left({}^\triangleright F^{X_1,X_2,M_1}_{M_2}\right)^{X_3;\alpha,\beta}_{M_3;\gamma,\delta}\adjincludegraphics[valign=c, scale=0.20]{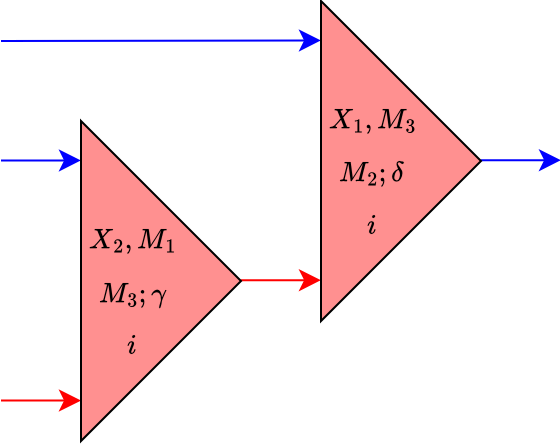},\label{2.37}
\end{equation}
which is derived from the associativity of the MPO actions on MPS \cite{MPO-SPT}. Note that we assume that it is independent to $i$. We also get
\begin{equation}
    \adjincludegraphics[valign=c, scale=0.20]{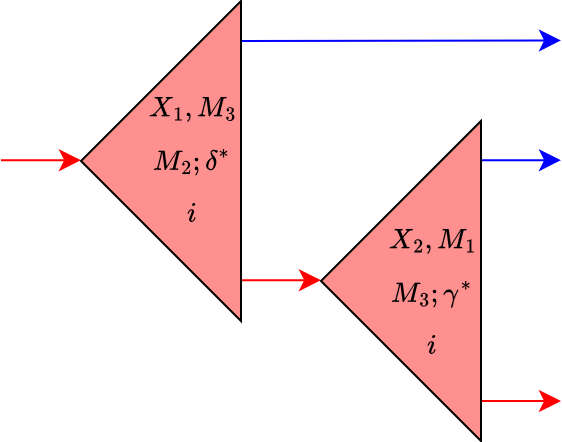}=\sum_{X_3}\sum_{\alpha,\beta}\left({}^\triangleright F^{X_1,X_2,M_1}_{M_2}\right)^{X_3;\alpha,\beta}_{M_3;\gamma,\delta}\adjincludegraphics[valign=c, scale=0.20]{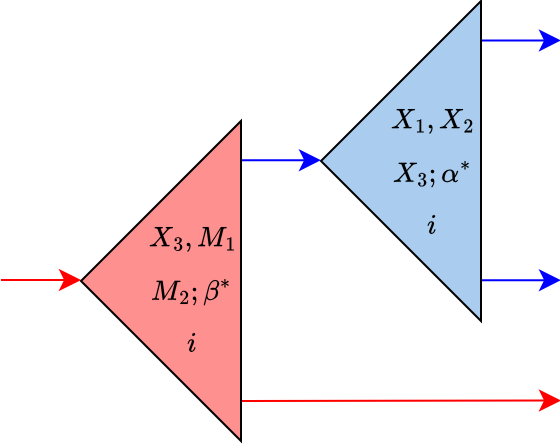}.\label{2.38}
\end{equation}

\textbf{Modulation:} The closedness condition of the lattice translation implies that the modulation of the modulated MPSs is described by an $F_T$-twisted $\mathcal{C}$-module autoequivalence $F_T^\mathcal{M}:\mathcal{M}\to\mathcal{M}$. The justification is as follows.

Suppose $\mathcal{M}$ corresponds to a phase. Given a ground state $M\in\mathcal{I}(\mathcal{M})$, it is sent to a different ground state $M'$ under lattice translation. This means that the $i+1$-th gate of $M$ is the $i$-th gate of $M'$ up to some unitary gates. We call the $i+1$-th gate of $M$ as the $i$-th gate of $F_T^\mathcal{M}M$
\begin{equation}
    \adjincludegraphics[valign=c, scale=0.20]{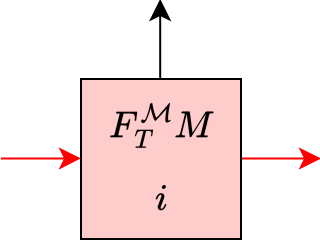}:=\adjincludegraphics[valign=c, scale=0.20]{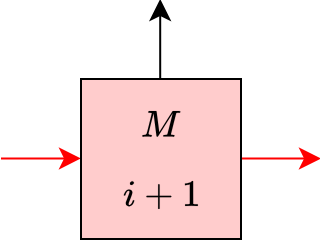}=\adjincludegraphics[valign=c, scale=0.20]{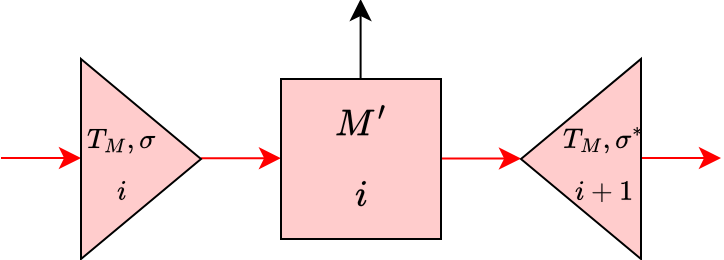},\label{2.39}
\end{equation}
$F_T^\mathcal{M}M$ is then an object in $\mathcal{M}$ satisfying $F_T^\mathcal{M}M\simeq M'$, and $\sigma$ is labeled by a unitary vector in the 1-dimensional vector space $\mathrm{Hom}_{\mathcal{M}}(F_T^\mathcal{M}M,M')$.

By definition, $F_T^\mathcal{M}$ is an endofunctor of $\mathcal{M}$ such that there exists a natural isomorphism
\begin{equation}
    \eta_T^\mathcal{M}:F_T^\mathcal{M}(X\triangleright M)\overset{\sim}\to F_T X\triangleright F_T^\mathcal{M} M.\label{2.40}
\end{equation}
Furthermore, we can construct another functor $F_{T^{-1}}^\mathcal{M}$ with respect to the inverse lattice translation $T^{-1}$ in the same way. $F_T^\mathcal{M}\circ F_{T^{-1}}^\mathcal{M}$ and $F_{T^{-1}}^\mathcal{M}\circ F_T^\mathcal{M}$ are equivalent to the identity functor $\id_\mathcal{M}$ by definition. Hence, $F_T^\mathcal{M}$ is an equivalent functor.

(\ref{2.37}) defines another isomorphism
\begin{equation}
    \begin{tikzcd}
    \mathrm{Hom}_{\mathcal{M}}(F_T^\mathcal{M}(X\triangleright M_1),M_2') \arrow[r,"\sim"] \arrow[d,"\vsim"]&\mathrm{Hom}_{\mathcal{M}}(X\triangleright M_1,M_2)\otimes\mathrm{Hom}(F_T^\mathcal{M}M_2,M_2') \arrow[d, "(\eta_T^{\mathcal{M}})^{X,M_1}_{M_2}"]\\
    \mathrm{Hom}_{\mathcal{M}}(F_TX\triangleright F_T^\mathcal{M}M_1,M_2') \arrow[r,"\sim"] &\mathrm{Hom}_{\mathcal{C}}(F_TX,X')\otimes\mathrm{Hom}_{\mathcal{M}}(F_T^\mathcal{M}M_1,M_1')\otimes\mathrm{Hom}_{\mathcal{M}}(X'\triangleright M_1',M_2')
\end{tikzcd}\label{2.41}
\end{equation}
which further defines the change of basis
\begin{equation}
    (\eta_T^{\mathcal{M}})^{X,M_1}_{M_2}(\alpha\otimes\sigma_2):=\sum_{\alpha'}\left((\eta_T^{\mathcal{M}})^{X,M_1}_{M_2}\right)^{\alpha,\sigma_2}_{\alpha',\tau,\sigma_1} \tau\otimes\sigma_1\otimes\alpha'\label{2.42}
\end{equation}
for $\alpha\in B^{X,M_1}_{M_2}$, $\alpha'\in B^{X',M_1'}_{M_2'}$, and unitary $\tau\in\mathrm{Hom}_{\mathcal{C}}(F_TX,X')$, $\sigma_a\in\mathrm{Hom}_{\mathcal{M}}(F_T^{\mathcal{M}}M_a,M_a')$ ($a=1,2$).

(\ref{2.42}) is captured by the tensor networks as follows. Acting the $i+1$-th gates of $X$ on $M_1$, we have the equation
\begin{align}
    \begin{split}
        &\sum_{M_2}\sum_{\alpha}\adjincludegraphics[valign=c, scale=0.20]{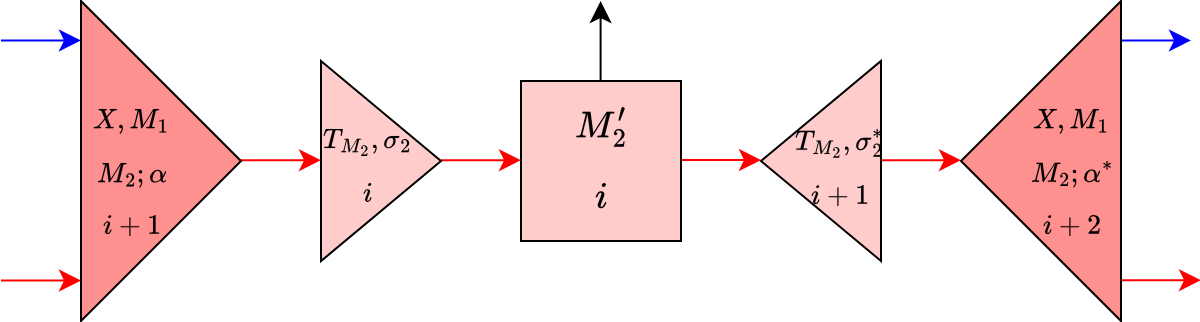}\\
        =&\sum_{M_2}\sum_{\alpha'}\adjincludegraphics[valign=c, scale=0.20]{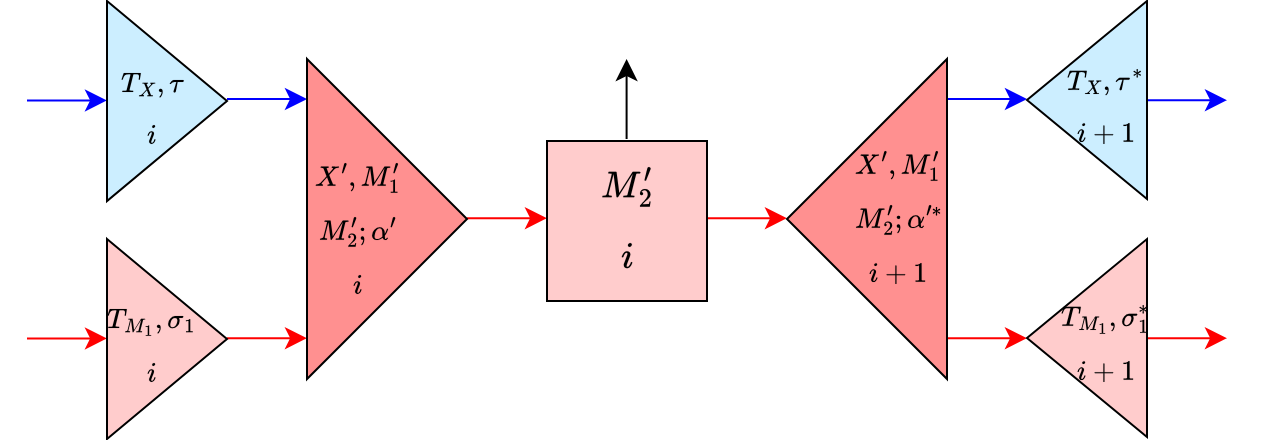}.
    \end{split}\label{2.43}
\end{align}
Define
\begin{equation}
    \left((\eta_T^{\mathcal{M}})^{X,M_1}_{M_2}\right)^{\alpha,\sigma_2}_{\alpha',\tau,\sigma_1}\adjincludegraphics[valign=c, scale=0.20]{2/MPS_3_2.png}:=\adjincludegraphics[valign=c, scale=0.20]{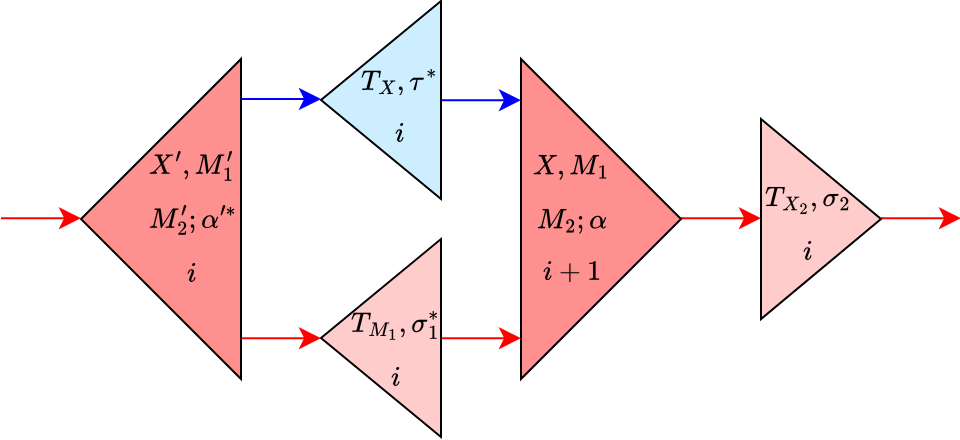},\label{2.44}
\end{equation}
and assume that it is independent to $i$. Then, due to (\ref{2.32}) and the injectivity of these MPSs, $\left((\eta_T^{\mathcal{M}})^{X,M_1}_{M_2}\right)^{\alpha,\sigma_2}_{\alpha',\tau,\sigma_1}$ is a complex number.

Lastly, according to (\ref{2.37}), (\ref{2.38}) and (\ref{2.44}), we have
\begin{align}
    \begin{split}
        &\adjincludegraphics[valign=c, scale=0.20]{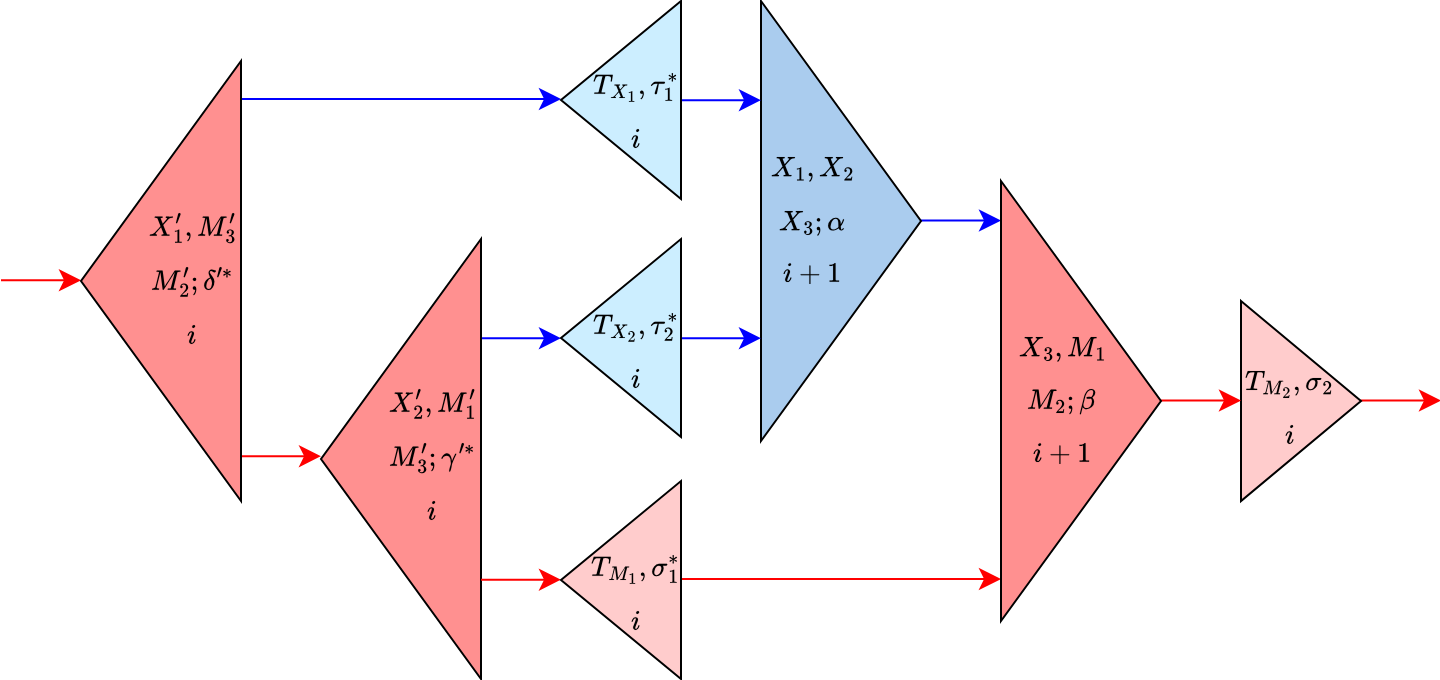}\\
        =&\sum_{M_3}\sum_{\gamma,\delta}\left((\eta_T^{\mathcal{M}})^{X_2,M_1}_{M_3}\right)^{\gamma,\sigma_3}_{\gamma',\tau_2,\sigma_1}\left((\eta_T^{\mathcal{M}})^{X_1,M_3}_{M_2}\right)^{\delta,\sigma_2}_{\delta',\tau_1,\sigma_3}\left({}^\triangleright F^{X_1,X_2,M_1}_{M_2}\right)^{X_3;\alpha,\beta}_{M_3;\gamma,\delta}\adjincludegraphics[valign=c, scale=0.20]{2/MPS_3_2.png}\\
        =&\sum_{X_3}\sum_{\alpha',\beta'}\left({}^\triangleright F^{X_1',X_2',M_1'}_{M_2'}\right)^{X_3';\alpha',\beta'}_{M_3';\gamma',\delta'}\left((\eta_T)^{X_1,X_2}_{X_3}\right)^{\alpha,\tau_3}_{\alpha',\tau_1,\tau_2}\left((\eta_T^{\mathcal{M}})^{X_3,M_1}_{M_2}\right)^{\beta,\sigma_2}_{\beta',\tau_3,\sigma_1}\adjincludegraphics[valign=c, scale=0.20]{2/MPS_3_2.png}.
    \end{split}\label{2.45}
\end{align}
This implies that the following diagram of isomorphisms commutes
\begin{equation}
    \begin{tikzcd}
        &{F_T(X_1\otimes X_2)\triangleright F_T^\mathcal{M}M}&\\
	{F_T^\mathcal{M}((X_1\otimes X_2)\triangleright M)} &&  {(F_TX_1\otimes F_TX_2)\triangleright F_T^\mathcal{M}M} \\
	&& {} \\
	{F_T^\mathcal{M}(X_1\triangleright(X_2\triangleright M))} &&  {F_TX_1\triangleright(F_TX_2\triangleright F_T^\mathcal{M}M)}\\
    &{F_TX_1\triangleright F_T^\mathcal{M}(X_2\triangleright M)}&
	\arrow["{\left(\eta_T^\mathcal{M}\right)_{X_1\otimes X_2, M}}", from=2-1, to=1-2]
	\arrow["{F_T^\mathcal{M}(m_{X_1,X_2,M})}"', from=2-1, to=4-1]
	\arrow["{(\eta_T)_{X_1,X_2}\otimes\mathrm{id}}", from=1-2, to=2-3]
	\arrow["{m_{F_TX_1,F_TX_2,F_T^\mathcal{M}M}}", from=2-3, to=4-3]
	\arrow["{\left(\eta_T^\mathcal{M}\right)_{X_1,X_2\triangleright M}}"', from=4-1, to=5-2]
	\arrow["{\mathrm{id}\otimes\left(\eta_T^\mathcal{M}\right)_{X_2,M}}"', from=5-2, to=4-3]
    \end{tikzcd}\label{2.46}
\end{equation}

Thus, $F_T^\mathcal{M}$ preserves the $F_T$-twisted $\mathcal{C}$-module structure. This concludes that $F_T^\mathcal{M}$ is an $F_T$-twisted $\mathcal{C}$-module autoequivalence of $\mathcal{M}$. 

Note that not every indecomposable semisimple $\mathcal{C}$-module category $\mathcal{M}$ equipped with such $F_T^\mathcal{M}$, this coincides with the results that not every SPT of the internal symmetry survived under modulation \cite{Dipolar SPT, Multipolar SPT, Modulated SPT}. Moreover, the choice of such $F_T^\mathcal{M}$ is not unique, and different choices correspond to different phases protected by both the internal symmetry and the lattice translation symmetry, known as the weak SPT \cite{Weak SPT, symmetry-enriched topological phases}. We will justify this statement by working on invertible modulated symmetries in the next subsection.

\subsection{Weak SPT Phases}\label{Subsec2.4}
The modulated SPT phases of invertible modulated symmetries have been studied in \cite{Modulated SPT}, and some specific models have been studied in \cite{Dipolar SPT, Multipolar SPT}. The framework in this work should recover these results.

Following the setup in Subsec. \ref{Subsec2.2}, let us classify the uniquely gapped phases. Since the ground states of a gapped phase $\mathcal{M}$ are labeled by the isomorphism classes of the simple objects in $\mathcal{M}$, we first fixed a representative $1$ of the simple objects, and define
\begin{equation}
    \psi(g_1,g_2):=\left({}^\triangleright F^{g_1,g_2,1}_{1}\right)^{g_1g_2;\alpha,\beta}_{1;\gamma,\delta}\in U(1)
\end{equation}
for fixed $B^{g,h}_{g,h}$ and $B^{g,1}_1$ ($g,h\in G$). The pentagon axiom gives the condition
\begin{equation}
    d\psi=-\omega,
\end{equation}
and hence, we can choose $B^{g,h}_{g,h}$ such that $\omega=0$. This implies that the condition of having uniquely gapped phases is $[\omega]=0\in H^3(G;U(1))$. Also, the redundancy of choosing $B^{g,1}_1$ gives the redundancy
\begin{equation}
    \psi\sim\psi+d\lambda
\end{equation}
for some $\lambda\in C^1(G;U(1))$. Therefore, the uniquely gapped phases are classified by $[\psi]\in H^2(G;U(1))$.

Now consider the modulation $F_T$ and $F_T^\mathcal{M}$. Define
\begin{equation}
    \rho(g):=\left((\eta_T^\mathcal{M})^{g,1}_1\right)^{\alpha,\sigma}_{\alpha',\tau_g,\sigma}
\end{equation}
for fixed $\tau_g:F_T g\to\phi_T(g)$ and $\sigma:F_T^\mathcal{M}1\to1$. (\ref{2.45}) then gives the condition
\begin{equation}
    d\rho=\phi_T^*\psi-\psi+\alpha.\label{2.51}
\end{equation}
According to (\ref{2.25}), we can choose $\tau_g$ such that $\alpha=0$. This is precisely why $\alpha$ is the mixed anomaly, since it captures the obstruction of having uniquely gapped phases. This further implies that the condition of $[\psi]\in H^2(G;U(1))$ describing a uniquely gapped phase is
\begin{equation}
    \phi_T^*[\psi]=[\psi].\label{2.52}
\end{equation}
This recovers the result (\ref{1.2}) shown in \cite{Modulated SPT}. Moreover, we have the redundancy
\begin{equation}
    \rho\sim\rho+\phi_T^*\lambda-\lambda.\label{2.53}
\end{equation}

We want to classify the pairs $(\rho,\psi)$ up to redundancy for a fixed $\phi_T\in\mathrm{Aut}(G)$. According to (\ref{2.52}), we have
\begin{equation}
    [\psi]\in H^2(G;U(1))^{\phi_T^*}:=\mathrm{ker}\left(1-\phi_T^*:H^2(G;U(1))\to H^2(G;U(1))\right).
\end{equation}
For a fixed $[\psi]$, according to (\ref{2.53}), the pairs $(\rho,\psi)$ up to redundancy form a $H^1(G;U(1))_{\phi_T^*}$-torsor, where
\begin{equation}
    H^1(G;U(1))_{\phi_T^*}:=\mathrm{coker}\left(1-\phi_T^*:H^1(G;U(1))\to H^1(G;U(1))\right).
\end{equation}
Therefore, the classification is described by the extension of $H^2(G;U(1))^{\phi_T^*}$ by $H^1(G;U(1))_{\phi_T^*}$.

Consider the case with no modulation, i.e., $\phi_T=\mathrm{id}_G$, then (\ref{2.51}) becomes 1-cocycle condition. Thus, the extension splits, and the classification is simply $H^1(G;U(1))\oplus H^2(G;U(1))$. The cohomology class $[\rho]\in H^1(G;U(1))$ actually captures the weak SPT phases\cite{Weak SPT, symmetry-enriched topological phases}, since different choices of $[\rho]$ are different phases only when they are also protected by the lattice translation symmetry.

For generic $\phi_T\in\mathrm{Aut}(G)$, the classification is described by the extension
\begin{equation}
    1\to H^1(G;U(1))_{\phi_T^*}\to H^2(G\underset{\phi}{\rtimes}\mathbb{Z};U(1))\to H^2(G;U(1))^{\phi_T^*}\to1.
\end{equation}
For detailed justification, please refer to Appendix \ref{A}.
\section{Lattice Translation Modulated 2+1D SymTFT}\label{Sec3}
We already have a well-established SymTFT description for uniform symmetries \cite{ICTP, Boundary-Bulk, Topological Holography, Symmetry as a shadow}, it is natural to ask how to build the bulk theories for modulated symmetries. In this section, we will give a general description of the bulk theories based on the results in the previous section.

\subsection{Modulated 2+1D SymTFT}\label{Subsec3.1}
Consider the 3-category of fusion categories $\mathbf{FusCat}$ constructed as follows.
\begin{itemize}
    \item 0-morphisms:
    \begin{equation}
        \mathrm{Ob}(\mathbf{FusCat}):=\{ \text{all fusion categories}\}.\label{3.1}
    \end{equation}
    \item 1-morphisms: Given $\mathcal{C},\mathcal{D}\in\mathrm{Ob}(\mathbf{FusCat})$,
    \begin{align}
        \begin{split}
            1\mathrm{Hom}_{\mathbf{FusCat}}(\mathcal{C},\mathcal{D}):=&\{\text{all finite semisimple }(\mathcal{C},\mathcal{D})\text{-bimodule categories}\}\\
            =&\{\text{all finite semisimple }(\mathcal{C}\boxtimes\mathcal{D}^{\mathrm{op}})\text{-module categories}\}.
        \end{split}\label{3.2}
    \end{align}
    \item Given $\mathcal{M}\in1\mathrm{Hom}_{\mathbf{FusCat}}(\mathcal{C},\mathcal{D})$ and $\mathcal{N}\in1\mathrm{Hom}_{\mathbf{FusCat}}(\mathcal{D},\mathcal{E})$, the 1-morphism composition is given by the Deligne tensor product over $\mathcal{D}$
    \begin{equation}
        \mathcal{N}\circ\mathcal{M}:=\mathcal{M}\boxtimes_{\mathcal{D}}\mathcal{N}\in1\mathrm{Hom}_{\mathbf{FusCat}}(\mathcal{C},\mathcal{E}).
    \end{equation}
    \item 2-morphisms: Given $\mathcal{M},\mathcal{N}\in1\mathrm{Hom}_{\mathbf{FusCat}}(\mathcal{C},\mathcal{D})$,
    \begin{equation}
        2\mathrm{Hom}_{\mathbf{FusCat}}(\mathcal{M},\mathcal{N}):=\mathrm{Fun}_{\mathcal{C}\boxtimes\mathcal{D}^{\mathrm{op}}}(\mathcal{M},\mathcal{N}).
    \end{equation}
    The 2-morphism composition is given by the $(\mathcal{C}\boxtimes\mathcal{D}^{\mathrm{op}})$-module functor composition.
    \item 3-morphisms: Given $F,G\in2\mathrm{Hom}_{\mathbf{FusCat}}(\mathcal{M},\mathcal{N})$,
    \begin{equation}
        3\mathrm{Hom}_{\mathbf{FusCat}}(F,G):=\{\text{all }(\mathcal{C}\boxtimes\mathcal{D}^{\mathrm{op}})\text{-module natural transformations}\}.
    \end{equation}
    The 3-morphism composition is given by the $(\mathcal{C}\boxtimes\mathcal{D}^{\mathrm{op}})$-module natural transformation composition.
\end{itemize}

Using these data, we can describe 3D TQFTs as follows \cite{Dualizable tensor categories}.

\begin{itemize}
    \item A 3D bulk is described by the Turaev-Viro model \cite{TV} of some $\mathcal{C}\in\mathrm{Ob}(\mathbf{FusCat})$, written as $TV(\mathcal{C})$. Since $TV(\mathcal{C})$ is equivalent to the Reshetikhin-Turaev model of the Drinfeld center $\mathcal{Z}(\mathcal{C})$ \cite{TV=RT, TV=RT1, TV=RT2}, the bulk is defined up to the equivalence
    \begin{equation}
        TV(\mathcal{C})\simeq TV(\mathcal{D})\iff \mathcal{Z}(\mathcal{C})\simeq\mathcal{Z}(\mathcal{D}).
    \end{equation}
    \item The vacuum is characterized by $TV(\mathrm{Vec})$, where $\mathrm{Vec}$ is the category of finite-dimensional $\mathbb{C}$-vector spaces.
    \item The 2D interface between two 3D bulks $TV(\mathcal{C})$ and $TV(\mathcal{D})$ is described by a 1-morphism in $1\mathrm{Hom}_{\mathbf{FusCat}}(\mathcal{C},\mathcal{D})$, which is a semisimple $(\mathcal{C}\boxtimes\mathcal{D}^{\mathrm{op}})$-module category. Thus, a 2D interface between $TV(\mathcal{C})$ and the vacuum (or usually called a boundary) is described by a semisimple $\mathcal{C}$-module category.
    \item If we choose a 2D boundary $\mathcal{M}\in1\mathrm{Hom}_{\mathbf{FusCat}}(\mathrm{Vec},\mathcal{D})$ to be indecomposable, then the Morita dual
    \begin{equation}
        \mathcal{D}^*_{\mathcal{M}}:=\mathrm{Fun}_{\mathcal{D}}(\mathcal{M},\mathcal{M})
    \end{equation}
    is also a fusion category \cite{Tensor Categories}. Moreover, $\mathcal{M}$ has a canonical $\mathcal{D}^*_{\mathcal{M}}$-module structure and $\mathcal{D}^*_{\mathcal{M}}$ is the symmetry of this boundary \cite{domain wall}.
\end{itemize}

In this work, we want to encode the modulation in the bulk. Given a modulated symmetry $(\mathcal{C},F_T)$ and a gapped phase $(\mathcal{M},F_T^{\mathcal{M}})$, denote the Morita dual as $\mathcal{D}:=\mathcal{C}^*_{\mathcal{M}}$. Suppose that there is a monoidal autoequivalence $F_T^{\mathcal{D}}:\mathcal{D}\to\mathcal{D}$ such that $F_T^{\mathcal{M}}$ is a $F_T^{\mathcal{D}}$-twisted $\mathcal{D}$-module autoequivalence, we should have
\begin{equation}
    F_T^{\mathcal{M}}(FM)\simeq \left(F_T^{\mathcal{D}}F\right)\left(F_T^{\mathcal{M}}M\right)
\end{equation}
for every $M\in\mathrm{Ob}(\mathcal{M})$ and $F\in\mathrm{Fun}_{\mathcal{C}}(\mathcal{M},\mathcal{M})$, and hence, 
\begin{equation}
    \left(F_T^{\mathcal{D}}F\right)\simeq F_T^{\mathcal{M}}\circ F\circ\left(F_T^{\mathcal{M}}\right)^{-1}.\label{3.9}
\end{equation}
Note that $\left(F_T^{\mathcal{M}}\right)^{-1}$ is uniquely defined up to equivalence. Since $F_T^{\mathcal{M}}$ is a $F_T$-twisted $\mathcal{C}$-module autoequivalence, we can always uniquely define such $F_T^{\mathcal{D}}$ satisfying (\ref{3.9}) up to equivalence. (\ref{3.9}) also implies that $F_T^{\mathcal{D}}$ is a monoidal autoequivalence. The modulated bulk that we aim to construct should be able to capture $F_T^{\mathcal{D}}$.

The idea is to consider a $\mathcal{D}$-bimodule category $\mathcal{T}:={}_{\mathrm{id}_{\mathcal{D}}}\mathcal{D}_{F_T^{\mathcal{D}}}$, which is $\mathcal{D}$ forgetting the monoidal structure and equipped with the right $\mathcal{D}$-module structure
\begin{align}
    \begin{split}
        \mathcal{T}\times\mathcal{D}&\to\mathcal{T},\\
        (T,Y)&\mapsto T\otimes F_T^{\mathcal{D}}Y
    \end{split}
\end{align}
and the regular left $\mathcal{D}$-module structure. By definition, $\mathcal{T}$ is invertible since $\mathcal{T}^{\mathrm{op}}\boxtimes_{\mathcal{D}}\mathcal{T}\simeq\mathcal{D}$ as $\mathcal{D}$-bimodule category.

\begin{figure}[H]
\centering
\includegraphics[scale = 0.12]{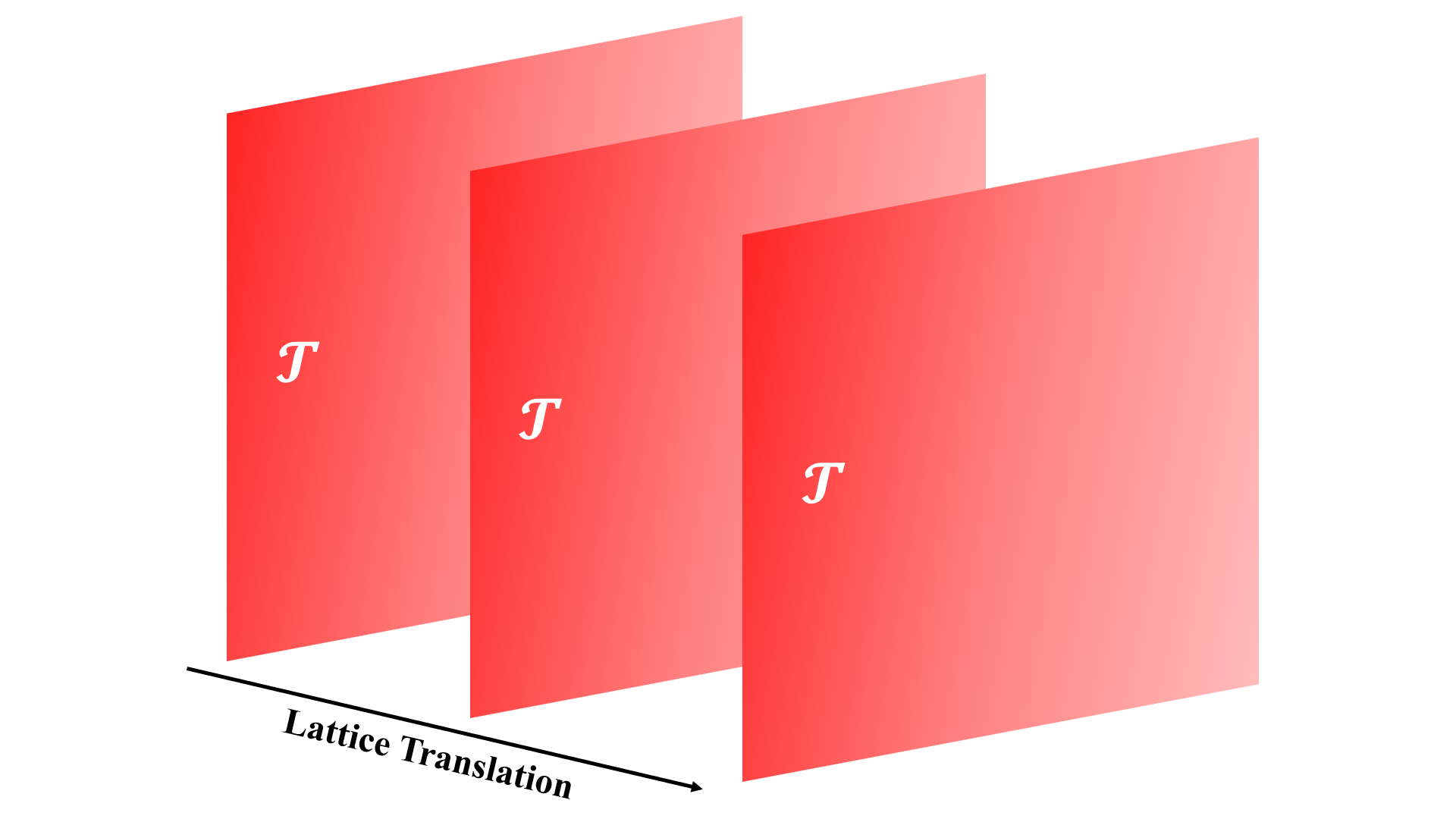}
\caption{Modulated TFT is constructed by inserting a series of domain walls described by an invertible bimodule category $\mathcal{T}$.}
\label{Modulated TFT}
\end{figure}

Why $\mathcal{T}$ is the structure we want? We can reinterpret $F_T^{\mathcal{M}}$ as a $\mathcal{D}$-module equivalence
\begin{equation}
    F_T^{\mathcal{M}}:\mathcal{M}\overset{\sim}{\to}\mathcal{M}_{F_T^{\mathcal{D}}},
\end{equation}
where $\mathcal{M}_{F_T^{\mathcal{D}}}$ is $\mathcal{M}$ with the new (right) $\mathcal{D}$-module structure given by
\begin{align}
    \begin{split}
        \mathcal{M}_{F_T^{\mathcal{D}}}\times\mathcal{D}&\to\mathcal{M}_{F_T^{\mathcal{D}}}.\\
        (M,Y)&\mapsto M\triangleleft F_T^{\mathcal{D}}Y
    \end{split}
\end{align}
Then, we have
\begin{equation}
    F_T^{\mathcal{M}}:\mathcal{M}\overset{\sim}{\to}\mathcal{M}_{F_T^{\mathcal{D}}}\simeq\mathcal{M}\boxtimes_{\mathcal{D}}\mathcal{T}.
\end{equation}
This can be captured by a 1D interface of two $\mathcal{M}$ and one $\mathcal{T}$. Thus, the modulated 3D bulk is described by $TV(\mathcal{D})$ with a series of domain walls, described by $\mathcal{T}$, inserted along the lattice translation direction. It is illustrated in Fig. \ref{Modulated TFT}.

There are several ways to build the modulated TFTs explicitly, based on the idea we propose in this subsection. One is directly inserts the domain walls in Turaev-Viro models. Since Levin-Wen models \cite{LW}
are equivalent to Turaev-Viro models \cite{LW=TV}, we will construct the modulated Levin-Wen model for $\mathbb{Z}_N$ dipole symmetry case in Subsec. \ref{Subsec3.3}; another one that we will discuss in Subsec. \ref{Subsec3.4} is considering BF theories with connection 1-form being modified by the presence of domain walls.

\subsection{Gapped Anyon Condensations}\label{Subsec3.2}
The correspondence between the boundary gapped phases and the bulk gapped anyon condensations has been well studied. The mathematical structure is given by \cite{DMNO10, Anyon condensation}, which states that there is a 1-to-1 correspondence
\begin{equation}
    \begin{tikzcd}
    \{\text{Lagrangian algebras in }\mathcal{Z}(\mathcal{D})\text{ up to Morita equivalence}\} \arrow[d, "\vsim"] & A \arrow[d, mapsto]\\
        \{\text{Indecomposable semisimple }\mathcal{D}\text{-module categories up to equivalence}\} & \mathcal{M}_A
\end{tikzcd}\label{3.14}
\end{equation}
where 
\begin{equation}
    \mathrm{Mod}_{\mathcal{Z}(\mathcal{D})}(A)\simeq\mathcal{D}^*_{\mathcal{M}_A}:=\mathrm{Fun}_{\mathcal{D}}(\mathcal{M}_A,\mathcal{M}_A).
\end{equation}
We can construct $\mathcal{M}_A$ as follows. The forgetful functor $\mathrm{Forg}:\mathcal{Z}(\mathcal{D})\to\mathcal{D}$ send $A$ to an algebra in $\mathcal{D}$, which is in general not even connected anymore. The $\mathcal{D}$-module category $\mathrm{Mod}_{\mathcal{D}}(\mathrm{Forg}A)$ is then not indecomposable. But, $\mathrm{Mod}_{\mathcal{D}}(\mathrm{Forg}A)$ is equivalent to the direct sum of copies of same indecomposable semisimple $\mathcal{D}$-module category. Such indecomposable semisimple $\mathcal{D}$-module category is the corresponding $\mathcal{M}_A$ \cite{DMNO10}.

For modulated cases in this work, we have to generalize this correspondence since the boundary gapped phases are classified by pairs of $(\mathcal{M},F_{\mathcal{T}}^{\mathcal{M}})$.

We first recall that there exists an isomorphism \cite{ENO10}
\begin{equation}
    \Phi:\mathrm{BrPic}(\mathcal{D})\overset{\sim}{\to}\mathrm{Aut}^{\mathrm{br}}_{\otimes}(\mathcal{Z}(\mathcal{D})),\label{3.16}
\end{equation}
where the right-hand side is the group of braided monoidal autoequivalence of $\mathcal{Z}(\mathcal{D})$ up to equivalence. The isomorphism is constructed as follows. Since $\mathcal{T}$ is an invertible $\mathcal{D}$-bimodule category, it is an indecomposable semisimple $\mathcal{D}$-module category. We then have \cite{Tensor Categories}
\begin{equation}
    \mathcal{Z}(\mathcal{D})\simeq(\mathcal{D}\boxtimes\mathcal{D}^*_{\mathcal{T}})^*_{\mathcal{T}}\simeq\mathcal{Z}(\mathcal{D}^*_{\mathcal{T}}).\label{3.17}
\end{equation}
The invertibility of $\mathcal{T}$ also implies the monoidal equivalence \cite{ENO10}
\begin{align}
    \begin{split}
        \mathcal{D}&\overset{\sim}{\to}\mathcal{D}^*_{\mathcal{T}}.\\
        X&\mapsto -\triangleleft X
    \end{split}\label{3.18}
\end{align}
Thus, we have a canonical $\Phi(\mathcal{T}):\mathcal{Z}(\mathcal{D})\overset{\sim}{\to}\mathcal{Z}(\mathcal{D}^*_{\mathcal{T}})\overset{\sim}{\to}\mathcal{Z}(\mathcal{D})$.

\begin{figure}[H]
\centering
\includegraphics[scale = 0.4]{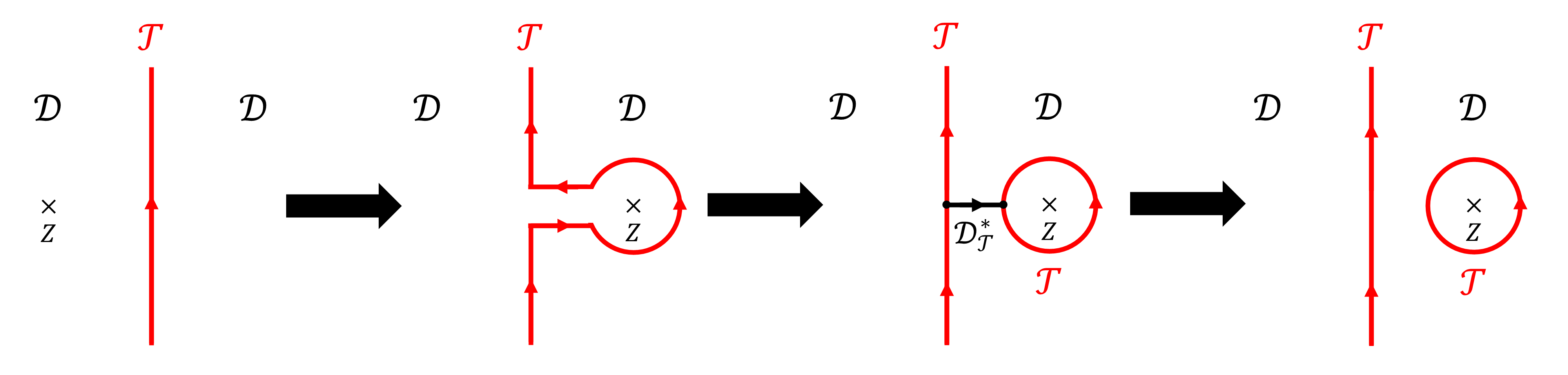}
\caption{An anyonic excitation $Z\in\mathcal{Z}(\mathcal{D})$ passes through the domain wall $\mathcal{T}$ shown in \cite{domain wall}.}
\label{Interface}
\end{figure}

The physical interpretation of (\ref{3.16}) is described in \cite{domain wall}, which states that an anyonic excitation $Z\in\mathcal{Z}(\mathcal{D})$ becomes $\Phi(\mathcal{T})Z$ when it passes through the domain wall $\mathcal{T}$ (along the lattice translation direction in our setup), and the monoidal structure and the braided structure are preserved. The graphical realization is illustrated in Fig. \ref{Interface}. Thus, according to (\ref{3.14}), a Lagrangian algebra $A$ corresponds to a gapped boundary if and only if $\Phi(\mathcal{T})A$ is Morita equivalent to $A$, i.e., 
\begin{equation}
    \mathrm{Mod}_{\mathcal{Z}(\mathcal{D})}(\Phi(\mathcal{T})A)\simeq\mathrm{Mod}_{\mathcal{Z}(\mathcal{D})}(A).\label{3.19}
\end{equation}

Note that a Lagrangian algebra $A$ satisfying (\ref{3.19}) only determines the corresponding indecomposable semisimple $\mathcal{D}$-module category $\mathcal{M}_A$, it does not fix the choices of $F_T^{\mathcal{D}}$-twisted $\mathcal{D}$-module autoequivalence of $\mathcal{M}_A$, which distinguish the weak SPT.
\subsection{Example: Modulated SymTFT of $\mathbb{Z}_N$ Dipole Symmetry}\label{Subsec3.3}
In this subsection, we consider the $\mathbb{Z}_N$ dipole symmetry as an example. The $\mathbb{Z}_N$ dipole symmetry on a 1+1D $\mathbb{Z}_N$ spin chain is generated by the symmetry operators
\begin{align}
    \begin{split}
        &U_Q:=\prod_{i}\sigma^x_i,\\
        &U_D:=\prod_i(\sigma^x_i)^i.
    \end{split}
\end{align}
Hence, the internal symmetry group is $\mathbb{Z}_N\times\mathbb{Z}_N$. Denote the symmetry operators as
\begin{equation}
    (Q,D):=(U_Q)^Q(U_D)^D=\prod_i(\sigma^x_i)^{Q+Di}.
\end{equation}
They are on-site by definition, and according to the definition of $F_T$, we have
\begin{equation}
    F_T(Q,D)=T^{-1}(Q,D)T=(Q+D,D).\label{3.23}
\end{equation}

Since $(\mathcal{C}^*_{\mathcal{M}})^*_{\mathcal{M}}\simeq\mathcal{C}$ for any indecomposable semisimple $\mathcal{C}$-module category $\mathcal{M}$, we want to build the bulk SymTFT a choice of Morita dual $\mathcal{C}^*_{\mathcal{M}}$. For a given modulated symmetry $(\mathcal{C},F_T)$, we can always choose the regular category $(\mathcal{C},F_T)$ as a phase. Here, we consider the Morita dual with respect to this regular choice of phase. We have an equivalence
\begin{align}
    \begin{split}
        \mathcal{D}:=\mathrm{Fun}_{\mathrm{Vec}_{\mathbb{Z}_N\times\mathbb{Z}_N}}(\mathrm{Vec}_{\mathbb{Z}_N\times\mathbb{Z}_N},\mathrm{Vec}_{\mathbb{Z}_N\times\mathbb{Z}_N})&\overset{\sim}{\to}\mathrm{Vec}_{\mathbb{Z}_N\times\mathbb{Z}_N}.\\
        \left((a,b)\mapsto(a+q,b+d)\right)&\mapsto(q,d)
    \end{split}\label{3.24}
\end{align}
According to (\ref{3.9}), the autoequivalence $F_T^{\mathcal{D}}$ of $\mathcal{D}$ should be
\begin{equation}
    F_T^{\mathcal{D}}(q,d)=(q+d,d).\label{3.25}
\end{equation}

Now, let us build the Hamiltonian of the modulated SymTFT. The idea is the same as the toric code on a square lattice (we can put it on a square lattice since the internal symmetry group is abelian), which is a stabilizer model with Levin-Wen model's ground states. For the modulated case, we also have to encode the data of the domain walls $\mathcal{T}\simeq{}_{\mathrm{id}}(\mathrm{Vec}_{\mathbb{Z}_N\times\mathbb{Z}_N})_{F_T^{\mathcal{D}}}$. 

We want to set the 1+1D boundary horizontally, so the domain walls are set on the vertical edges. For every horizontal edge coming from the left, its effect on the domain wall is modified by $F_T^{\mathcal{D}}$. Put two $\mathbb{Z}_N$ spins on each edge, and consider two sets of Pauli matrices $Z^Q,X^Q$ and $Z^D,X^D$ acting on these two spins, respectively. Graphically, we draw $Z^Q,X^Q$ in blue and $Z^D,X^D$ in red. The flatness condition of the ground states written in $Z^Q$ and $Z^D$ basis becomes
\begin{equation}
    \adjincludegraphics[valign=c, scale=0.25]{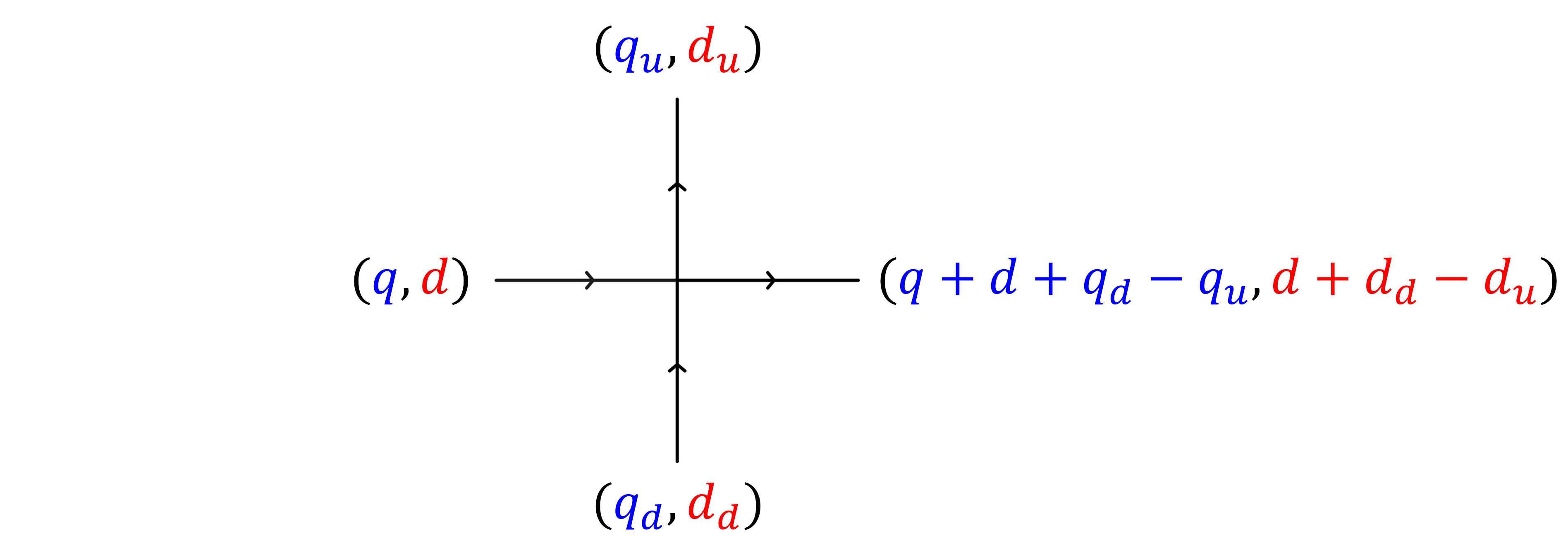}=1,\label{3.26}
\end{equation}
and the loop
\begin{equation}
    \adjincludegraphics[valign=c, scale=0.25]{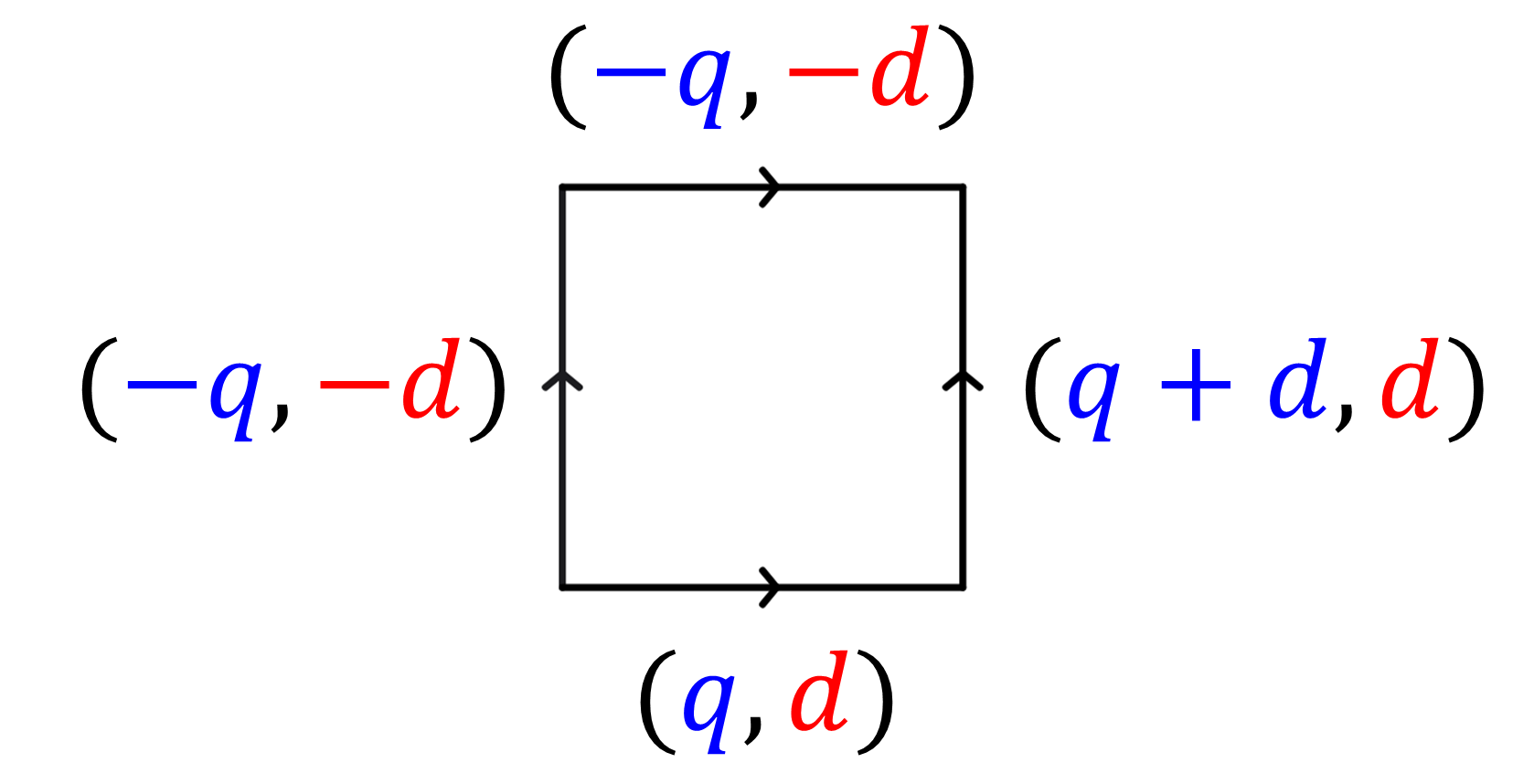}\label{3.27}
\end{equation}
is contractable. 

\begin{figure}[H]
\centering
\includegraphics[scale = 0.3]{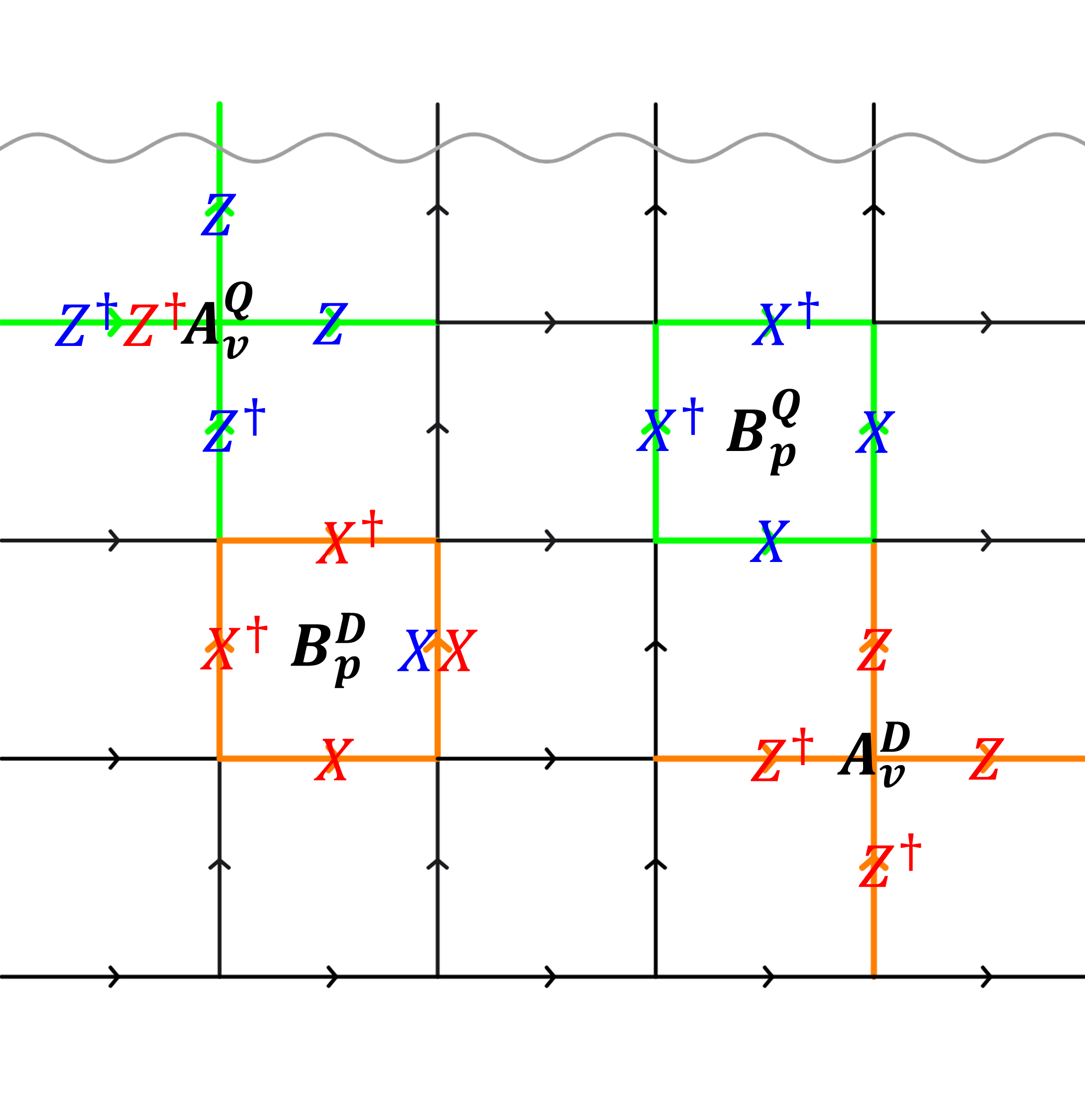}
\caption{Stabilizers of the modulated bulk model.}
\label{SymTFT Stabilizers}
\end{figure}

The condition (\ref{3.26}) defines the stabilizers $A_v^Q$ and $A_v^D$ shown in Fig. \ref{SymTFT Stabilizers}; the condition (\ref{3.27}) defines the stabilizers $B_p^Q$ and $B_p^D$ shown in Fig. \ref{SymTFT Stabilizers}. The bulk is then the stabilizer model
\begin{equation}
    H_{bulk}:=-\sum_{a=1}^N\left\{\sum_v\left[\left(A_v^Q\right)^a+\left(A_v^D\right)^a\right]+\sum_p\left[\left(B_p^Q\right)^a+\left(B_p^D\right)^a\right]\right\}.
\end{equation}

\begin{figure}[H]
\begin{subfigure}[t]{0.5\textwidth}
\centering
\includegraphics[scale = 0.3]{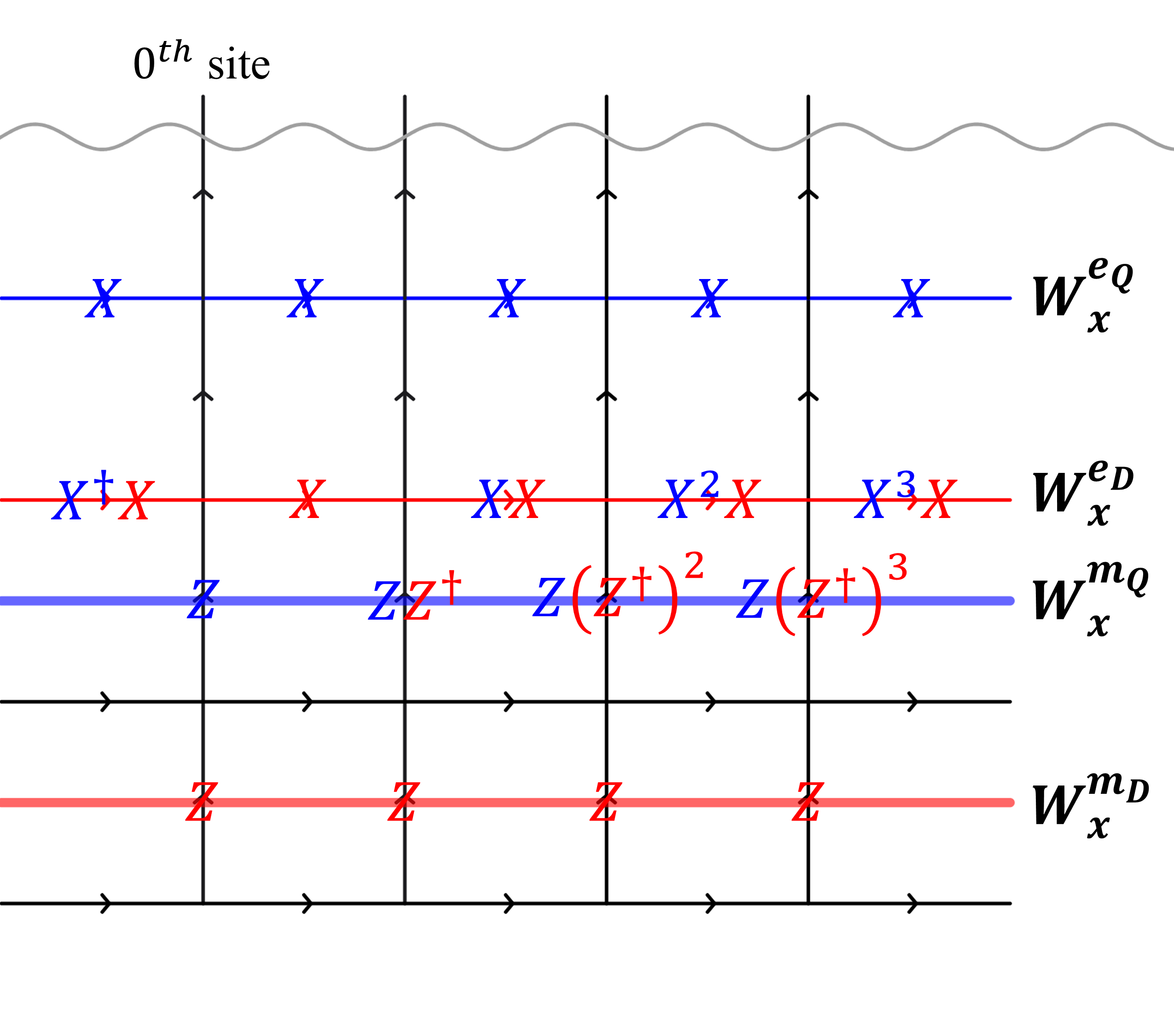}
\caption{}
\label{SymTFT X Operators}
\end{subfigure}
\begin{subfigure}[t]{0.5\textwidth}
\centering
\includegraphics[scale = 0.3]{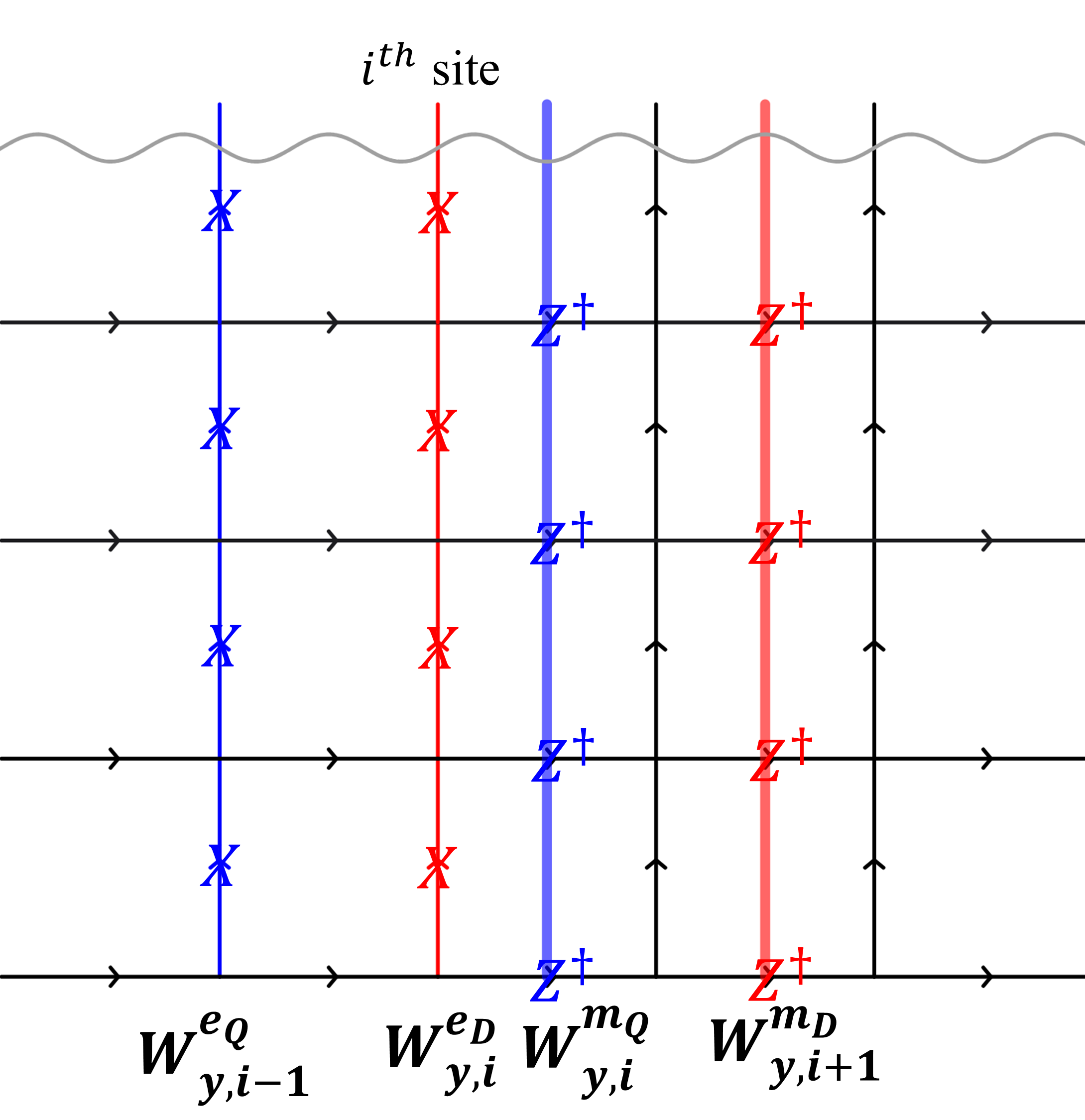}
\caption{}
\label{SymTFT Y Operators}
\end{subfigure}
\caption{(a) The anyonic loops that correspond to the symmetries/twists in the SymTFT picture. (b) The anyons that correspond to charges/fluxes in the SymTFT picture.}
\label{SymTFT Operators}
\end{figure}

The simple objects of the Drinfeld center $\mathcal{Z}(\mathrm{Vec}_{\mathbb{Z}_N\times\mathbb{Z}_N})$ are labeled by $((a,b),(\gamma,\delta))$, where $a,b\in\mathbb{Z}_N$ and $\gamma, \delta\in\widehat{\mathbb{Z}_N}\simeq\mathbb{Z}_N$. They can be generated by the anyons $e_Q:=((1,0),(\underline{0},\underline{0})), m_Q:=((0,0),(\underline{1},\underline{0})), e_D:=((0,1),(\underline{0},\underline{0})), m_D:=((0,0),(\underline{0},\underline{1}))$, shown in Fig. \ref{SymTFT Operators}. We can see that, under lattice translation
\begin{align}
    \begin{split}
        e_Q&\mapsto e_Q,\\
        e_D&\mapsto e_Qe_D,\\
        m_Q&\mapsto {m_Q}^{-1}m_D,\\
        m_D&\mapsto m_D,
    \end{split}\label{3.29}
\end{align}
which is exactly the image of $\mathcal{T}$ under the isomorphism (\ref{3.16}).

Let us now consider the effective 1+1D sandwich. We consider the bottom boundary to be smooth and study different choices of anyonic condensations on the top boundary. If we want the anyon condensations to be consistent, it is obvious that the condensable algebras should be closed under the lattice translation. This is exactly the statement (\ref{3.19}).

Consider first the case of the smooth top boundary, i.e., $m_Q$ and $m_D$ are condensed. In the effective 1+1D sandwich, $W_{y,i}^{m_Q}$ and $W_{y,i}^{m_D}$ can be identified as the $\mathbb{Z}_N$ charges ${\sigma^z_i}^\dagger$ and ${\tau^z_i}^\dagger$, respectively. According to the commutation relations, we have $W_x^{e_Q}=\prod_i\sigma^x_i$ and $W_x^{e_D}=\prod_i\left(\sigma^x_i\right)^i\tau^x_i$, which generate the internal symmetry of the effective 1+1D sandwich. The local operators on the bottom boundary then have the following forms
\begin{align}
    \begin{split}
            \adjincludegraphics[valign=c, scale=0.3]{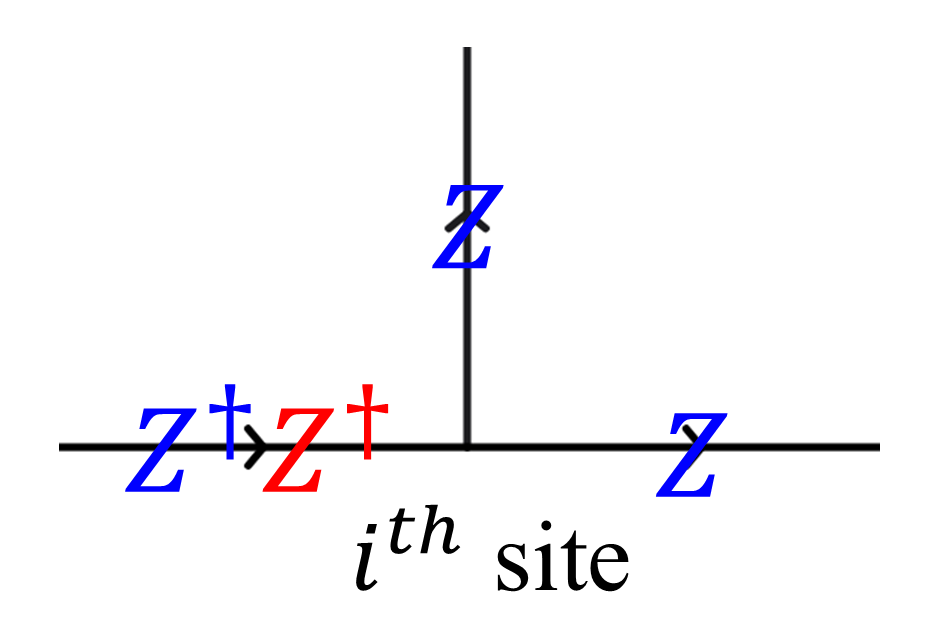}&=W_{y,i-1}^{m_Q}{W_{y,i-1}^{m_D}}{W_{y,i}^{m_Q}}^\dagger={\sigma^z_{i-1}}^\dagger{\tau^z_{i-1}}^\dagger\sigma^z_i,\\
            \adjincludegraphics[valign=c, scale=0.3]{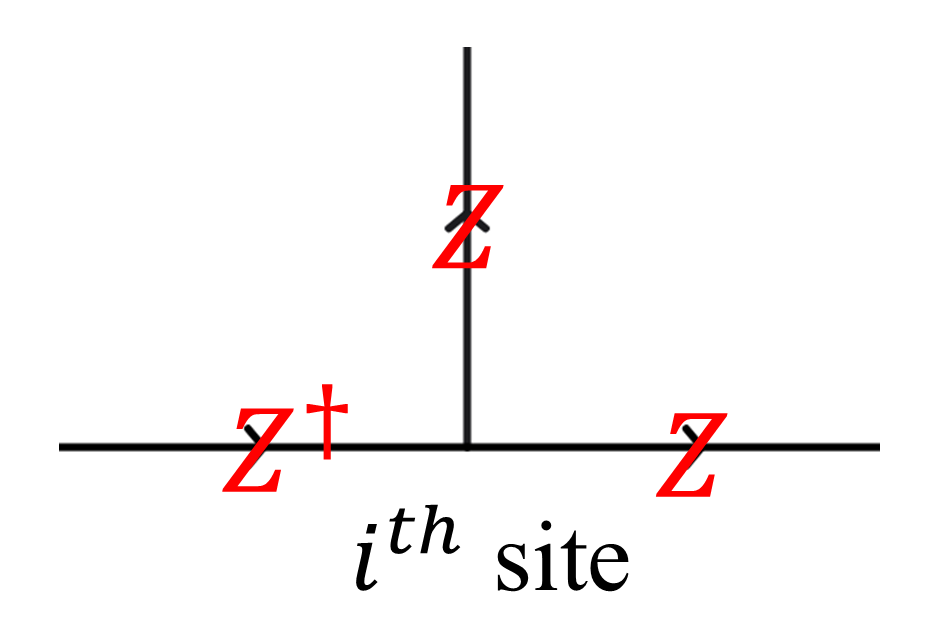}&=W_{y,i-1}^{m_D}{W_{y,i}^{m_D}}^\dagger={\tau^z_{i-1}}^\dagger\tau^z_i,\\
            \adjincludegraphics[valign=c, scale=0.3]{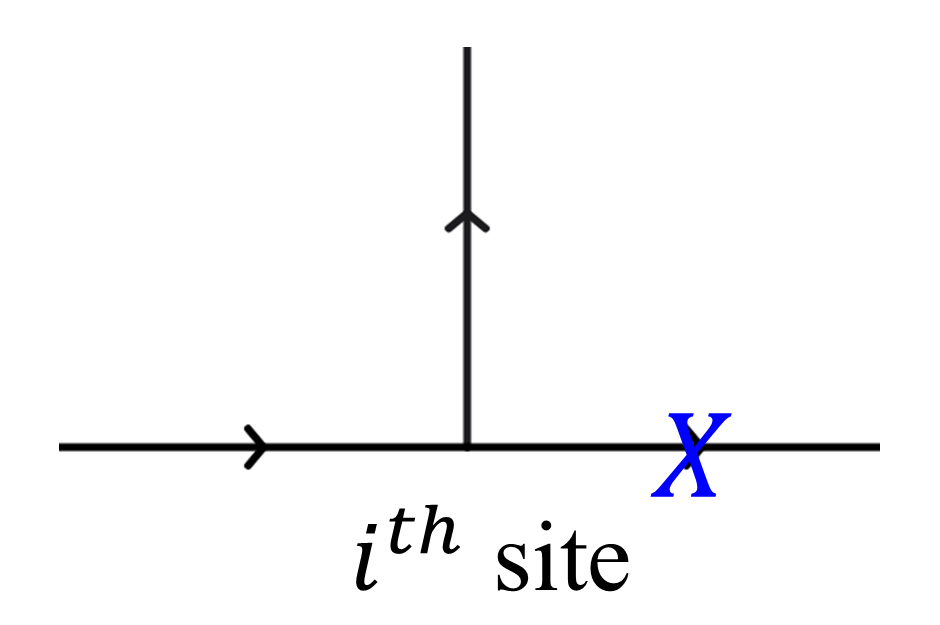}&=\sigma^x_i,\\
            \adjincludegraphics[valign=c, scale=0.3]{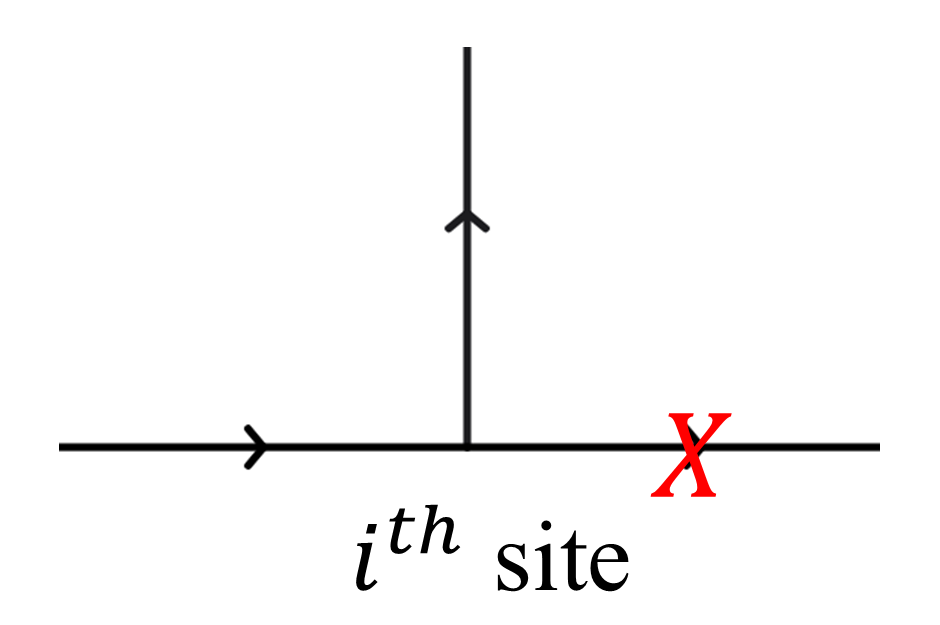}&=\tau^x_i.
    \end{split}
\end{align}
We can further identify $W_{y,i}^{e_Q}$ and $W_{y,i}^{e_D}$ as $\mathbb{Z}_N$ fluxes, and we can get the twisted Hamiltonian for different flux sectors.

Let us take a closer look at the symmetry operators of the phase. The corresponding indecomposable semisimple $\mathcal{D}$-module category of this Lagrangian algebra is the regular $\mathcal{D}$-module category $\mathcal{M}=\mathrm{Vec}_{\mathbb{Z}_N\times\mathbb{Z}_N}$. According to the discussion in Subsec. \ref{Subsec2.4}, the only possible $F_T^{\mathcal{D}}$-twisted $\mathcal{D}$-module autoequivalence $F_T^{\mathcal{M}}$ is simply $F_T^{\mathcal{D}}$. Then, according to the definition of $\mathcal{D}$ and $F_T^\mathcal{D}$ we use in this subsection, this modulated symmetry of this phase is apparently the original $(\mathcal{C},F_T)$.

The second case is the rough top boundary, i.e., $e_Q$ and $e_D$ are condensed. Similarly, $W_{y,i}^{e_Q}$ and $W_{y,i}^{e_D}$ can be identified as the $\mathbb{Z}_N$ charges ${\sigma^z_i}^\dagger$ and ${\tau^z_i}^\dagger$, respectively. According to the commutation relations, we have $W_x^{m_Q}=\prod_i\sigma^x_i\left({\tau^x_i}^\dagger\right)^i$ and $W_x^{m_D}=\prod_i\tau^x_i$, which generate the internal symmetry of the effective 1+1D sandwich. The local operators on the bottom boundary then have the following forms
\begin{align}
    \begin{split}
            \adjincludegraphics[valign=c, scale=0.3]{3/SymTFT_Local_X_Q.png}&=W_{y,i}^{e_Q}{W_{y,i+1}^{e_D}}^\dagger={\sigma^z_i}^\dagger\sigma^z_{i+1},\\
            \adjincludegraphics[valign=c, scale=0.3]{3/SymTFT_Local_X_D.png}&=W_{y,i}^{e_D}{W_{y,i+1}^{e_Q}}^\dagger{W_{y,i+1}^{e_D}}^\dagger={\tau^z_i}^\dagger\sigma^z_{i+1}\tau^z_{i+1},\\
            \adjincludegraphics[valign=c, scale=0.3]{3/SymTFT_Local_Z_Q.png}&=\sigma^x_i,\\
            \adjincludegraphics[valign=c, scale=0.3]{3/SymTFT_Local_Z_D.png}&=\tau^x_i.
    \end{split}
\end{align}
We can further identify $W_{y,i}^{m_Q}$ and $W_{y,i}^{m_D}$ as $\mathbb{Z}_N$ fluxes, and we can get the twisted Hamiltonian for different flux sectors.

One important thing we can observe is that, the roles of $\sigma$ and $\tau$ as monopole and dipole interchange. The reason is as follows.

The corresponding indecomposable semisimple $\mathcal{D}$-module category of this Lagrangian algebra is $\mathcal{M}=\mathrm{Vec}$. Fix a set of $[g]\in\mathrm{Hom}_{\mathrm{Vec}}(g\triangleright1,1)$ for every $g\in\mathbb{Z}_N\times\mathbb{Z}_N$, such that all the ${}^\triangleright F$-symbols are trivial. The internal symmetry then has the form
\begin{align}
    \begin{split}
        \mathrm{Rep}(\mathbb{Z}_N\times\mathbb{Z}_N)&\overset{\sim}{\to}\mathrm{Fun}_{\mathbb{Z}_N\times\mathbb{Z}_N}(\mathrm{Vec},\mathrm{Vec}).\\
        \alpha&\mapsto \left([g]\mapsto\alpha(g)[g]\right)=:F_\alpha
    \end{split}
\end{align}
On the other hand, according to the discussion in Subsec. \ref{Subsec2.4}, $F_T^{\mathcal{M}}$ in general has the form
\begin{equation}
    F_T^{\mathcal{M}}([g])=\beta(g)\left[\phi(g)\right]
\end{equation}
for some $\beta\in\widehat{\mathbb{Z}_N\times\mathbb{Z}_N}\simeq H^1(\mathbb{Z}_N\times\mathbb{Z}_N)$, where $\phi(q,d):=F_T^{\mathcal{D}}(q,d)=(q+d,d)$. The modulation of the symmetry is given by $F_\alpha\mapsto F_T^{\mathcal{M}}\circ F_\alpha\circ\left(F_T^{\mathcal{M}}\right)^{-1}$, where
\begin{align}
    \begin{split}
        F_T^{\mathcal{M}}\circ F_\alpha\circ\left(F_T^{\mathcal{M}}\right)^{-1}[g]&=\beta\left(\phi^{-1}(g)\right)^{-1}F_T^{\mathcal{M}}\circ F_\alpha\left[\phi^{-1}(g)\right]\\
        &=\alpha\left(\phi^{-1}(g)\right)
        \beta\left(\phi^{-1}(g)\right)^{-1}F_T^{\mathcal{M}}\left[\phi^{-1}(g)\right]=\alpha\left(\phi^{-1}(g)\right)[g]\\
        &=F_{\alpha\circ\phi^{-1}}[g].
    \end{split}
\end{align}
Lastly, consider the equivalence
\begin{align}
    \begin{split}
        \mathrm{Vec}_{\mathbb{Z}_N\times\mathbb{Z}_N}&\overset{\sim}{\to}\mathrm{Rep}(\mathbb{Z}_N\times\mathbb{Z}_N).\\
        (Q,D)&\mapsto \left((a,b)\mapsto e^{\frac{2\pi i}{N}(aQ+bD)}\right)
    \end{split}
\end{align}
Therefore, the modulation of the symmetry is
\begin{align}
    \mathrm{Vec}_{\mathbb{Z}_N\times\mathbb{Z}_N}&\overset{\sim}{\to}\mathrm{Vec}_{\mathbb{Z}_N\times\mathbb{Z}_N},\\
    (Q,D)&\mapsto(Q,D-Q)
\end{align}
which is exactly what we get from the effective 1+1D sandwich.

\subsection{Foliated BF Theory}\label{Subsec3.4}
In this subsection, we will demonstrate that the UV limit of our construction for abelian cases meets the construction of foliated BF theory. We will use the $\mathbb{Z}_N$ dipole symmetry discussed in the previous subsection as an example.

Since the internal symmetry group is $\mathbb{Z}_N\times\mathbb{Z}_N$, we first write down a $\mathbb{Z}_N\times\mathbb{Z}_N$ BF theory
\begin{equation}
    \mathcal{L}_{\mathbb{Z}_N\times\mathbb{Z}_N}=\frac{N}{2\pi}\left(b^Q\wedge da^Q+b^D\wedge da^D\right).
\end{equation}

Due to the presence of domain walls, we should add an additional term given by the modulation (\ref{3.23})
\begin{equation}
    a^Q_\mu(t,x+1,y)-a^Q_\mu(t,x,y)\mapsto a^Q_\mu(t,x+1,y)-\left(a^Q_\mu(t,x,y)+a^D_\mu(t,x,y)\right),
\end{equation}
and the continuum limit is
\begin{align}
    \begin{split}
        \partial_xa^Q_\mu&\mapsto\partial_xa^Q_\mu-a^D_\mu,\\
        \implies da^Q&\mapsto da^Q+a^D\wedge dx.
    \end{split}
\end{align}

Thus, the Lagrangian of the modulated bulk of the dipole symmetry is
\begin{align}
    \begin{split}
        \mathcal{L}_{\mathrm{dip}}&=\frac{N}{2\pi}\left(b^Q\wedge\left(da^Q+a^D\wedge dx\right)+b^D\wedge da^D\right)\\
        &=\frac{N}{2\pi}\left(b^Q\wedge da^Q+b^D\wedge da^D+b^Q\wedge a^D\wedge dx\right).
    \end{split}
\end{align}
This Lagrangian is exactly the foliated BF theory for dipole symmetry. Similar to the construction of foliated BF theory shown in \cite{Foliated BF}, but we only consider the dipole symmetry along x-axis. The construction is as follows.

Write the symmetry actions $Q$ and $D$ as
\begin{align}
    &Q=\int_V *j^Q,\\
    &D=\int_V *J^D,
\end{align}
where $V$ is an equal-time codimension-1 submanifold. Since $j^Q$ and $J^D$ are conserved currents, we have
\begin{align}
    &d*j^Q=0,\label{3.38}\\
    &d*J^D=0.
\end{align}

According to (\ref{3.25}), we have
\begin{align}
    &Q(x+1,y)=Q(x,y),\\
    &D(x+1,y)=D(x,y)+Q(x,y).
\end{align}
The continuum limit is
\begin{align}
    &\partial_i Q=0,\\
    &\partial_i D=\delta_{x,i}Q.
\end{align}
Hence, we can write
\begin{equation}
    *J^D=*j^D+x*j^Q,
\end{equation}
where $j^D$ is defined to be local.

Couple the currents $j^Q,j^D$ to the background 1-form gauge fields $a^Q,a^D$
\begin{equation}
    S=\int_M\left(a^Q\wedge*j^Q+a^D\wedge*j^D\right).
\end{equation}
According to (\ref{3.38}) and
\begin{equation}
    d*j^D=-dx\wedge*j^Q,
\end{equation}
the gauge symmetry is then
\begin{align}
    a^Q&\mapsto a^Q+d\lambda^Q-\lambda^D dx,\\
    a^D&\mapsto a^D+d\lambda^D.
\end{align}
Thus, the gauge invariant 2-forms are
\begin{align}
    &f^Q:=da^Q+a^D\wedge dx,\\
    &f^D:=da^D,
\end{align}
and the Lagrangian of the foliated BF theory is
\begin{align}
    \begin{split}
        \mathcal{L}&=\frac{N}{2\pi}\left(b^Q\wedge f^Q+b^D\wedge f^D\right)\\
        &=\frac{N}{2\pi}\left(b^Q\wedge da^Q+b^D\wedge da^D+b^Q\wedge a^D\wedge dx\right).
    \end{split}
\end{align}

Therefore, our construction recovers the foliated BF theory.
\section{Summary and Outlooks}
We have proposed a natural way to describe the algebraic structure of the lattice translation modulated symmetries in 1+1D, and classified the corresponding SPTs. The modulation of the internal symmetry $\mathcal{C}$ is described by a monoidal autoequivalence $F_T:\mathcal{C}\to\mathcal{C}$, which also captures the mixed anomaly of the internal symmetry and the lattice translation. The modulation of the $\mathcal{C}$-SPT $\mathcal{M}$ is described by an $F_T$-twisted $\mathcal{C}$-module autoequivalence $F^{\mathcal{M}}_T:\mathcal{M}\to\mathcal{M}$, and different choices of such $F^{\mathcal{M}}_T$ classify the weak SPT. These structures recover some known results on invertible modulated symmetries, and provide a general framework for describing non-invertible modulated symmetries. The construction of the modulated version of the 2+1D SymTFT bulk naturally arises from all these autoequivalences on the boundary, which is inserting a series of domain walls, described by the corresponding invertible bimodule category, along the lattice translation direction. The anyons in the bulk will then alter along the lattice translation direction due to the presence of the domain walls.

There are several future research directions that we would like to study. The most straightforward one is to study more specific models, especially the ones with non-trivial mixed anomalies or non-invertible modulated symmetries. In addition, in this work we only focus on the lattice translation modulation in 1+1D, it is natural to generalize the discussion for different types of modulations. Another future research direction is to generalize the modulated symmetries to higher dimensional and higher-form cases.

\section*{Acknowledgement}
CYY acknowledges helpful comments and suggestions from Kansei Inamura, Kantaro Ohmori, Masahito Yamazaki and Tsubasa Oishi. CYY thanks Kansei Inamura and Myles Scollon for comments on the manuscript. This work was inspired in part during the Theoretical Sciences Visiting Program (TP25QM) at the Okinawa Institute of Science and Technology. CYY is supported by the GSGC Program, University of Tokyo.
\begin{appendices}
    \section{Group Cohomology of Semidirect Products}\label{A}
Consider the semidirect product $E:=G\underset{\phi}{\rtimes}\mathbb{Z}$ discussed in Subsec. \ref{Subsec2.2} and Subsec. \ref{Subsec2.4}. We have an extension
\begin{equation}
    1\to G\overset{i}{\to} E\to \mathbb{Z}\to 1.
\end{equation}

We can calculate $H^k(E)$ by studying the second page of the Lyndon-Hochschild-Serre spectral sequence. Since $H^p(\mathbb{Z},-)=0$ for $p>1$, we have the short exact sequence
\begin{equation}
    1\to E^{1,k-1}_2\to H^k(E)\to E^{0,k}_2\to1,
\end{equation}
where
\begin{equation}
    E^{p,q}_2=H^p(\mathbb{Z};H^q(G)).
\end{equation}

The $\mathbb{Z}$-action on $H^q(G)$ is given by
\begin{align}
    \begin{split}
        \mathbb{Z}&\to \mathrm{End}(H^q(G)),\\
        a&\mapsto \phi(a)^*=(\phi_T^*)^a
    \end{split}
\end{align}
where $\phi_T:=\phi(1)$.

Since $H^q(G)$ are non-trivial $\mathbb{Z}$-modules, we don't have $H^0(\mathbb{Z};H^q(G))\simeq H^q(G) \simeq H^1(\mathbb{Z};H^q(G))$. Instead, we get
\begin{equation}
    H^0(\mathbb{Z};H^q(G))\simeq H^q(G)^{\phi_T^*}:=\mathrm{ker}\left(1-\phi_T^*:H^q(G)\to H^q(G)\right), 
\end{equation}
which is the invariance of $H^q(G)$ with respect to $\phi_T^*$, and
\begin{equation}
    H^1(\mathbb{Z};H^q(G))\simeq H^q(G)_{\phi_T^*}:=\mathrm{coker}\left(1-\phi_T^*:H^q(G)\to H^q(G)\right), 
\end{equation}
which is the coinvariance of $H^q(G)$ with respect to $\phi_T^*$. Therfore, we can rewrite the short exact sequence as
\begin{equation}
    1\to H^{k-1}(G)_{\phi_T^*}\to H^k(E)\to H^k(G)^{\phi_T^*}\to1
\end{equation}

Note that we can also obtain this short exact sequence by the Wang exact sequence of the fibration $BG\to BE\to B\mathbb{Z}\simeq S^1$
\begin{equation}
    \cdots\to H^{k-1}(G)\overset{1-\phi_T^*}{\to}H^{k-1}(G)\overset{\delta^{k-1}}{\to}H^k(E)\overset{i^*}{\to}H^k(G)\overset{1-\phi_T^*}{\to}H^k(G)\to\cdots.
\end{equation}

We want to study whether this extension is the same as the one we have in Subsec. \ref{Subsec2.2} and Subsec. \ref{Subsec2.4}. To do so, let us write down the double complex
\begin{equation}
    \begin{tikzcd}
        \vdots &\vdots \\
        C^*(G) \arrow[r, "1-\phi_T^*"] \arrow[u, "d"] & C^*(G) \arrow[u, "d"]\\
        C^{*-1}(G) \arrow[r, "1-\phi_T^*"] \arrow[u, "d"] & C^{*-1}(G) \arrow[u, "d"]\\
        \vdots \arrow[u, "d"] &\vdots \arrow[u, "d"]
\end{tikzcd}
\end{equation}
which only has two columns since $\mathbb{Z}$ has cohomological dimension 1. The group cohomology of $E$ is the cohomology of the total complex of this double complex. Therefore, $H^k(E)$ is isomorphic to $(\alpha,\omega)\in C^{k-1}(G)\oplus C^k(G)$ satisfying
\begin{equation}
    d\omega=0,
\end{equation}
\begin{equation}
    d\alpha+(-1)^k(1-\phi_T^*)\omega=0,
\end{equation}
and quotient by the equivalence
\begin{equation}
    \omega\sim\omega+d\beta,
\end{equation}
\begin{equation}
    \alpha\sim\alpha+d\gamma+(-1)^{k-1}(1-\phi_T^*)\beta,
\end{equation}
for $(\gamma,\beta)\in C^{k-2}(G)\oplus C^{k-1}(G)$. Rewrite $\alpha$ and $\gamma$ as $(-1)^k\alpha$ and $(-1)^k\gamma$, then we get the exactly same classifications in Subsec. \ref{Subsec2.2} and Subsec. \ref{Subsec2.4}. Therefore, the anomaly of a modulated symmetry $\phi:\mathbb{Z}\to\mathrm{Aut}(G)$ is captured by $H^3(G\underset{\phi}{\rtimes}\mathbb{Z};U(1))$, and the uniquely gapped modulated SPTs are classified by $H^2(G\underset{\phi}{\rtimes}\mathbb{Z};U(1))$ when the symmetry is anomaly-free.

\end{appendices}

\end{document}